\documentclass{egpubl}
\usepackage{pg2023s}

\WsShortPaper      

\usepackage[T1]{fontenc}
\usepackage{dfadobe}  

\usepackage{cite}  
\BibtexOrBiblatex
\electronicVersion
\PrintedOrElectronic
\ifpdf \usepackage[pdftex]{graphicx} \pdfcompresslevel=9
\else \usepackage[dvips]{graphicx} \fi

\usepackage{booktabs} 
\usepackage{multirow}
\usepackage{caption}
\usepackage{subcaption}
\usepackage{hyperref}
\usepackage{egweblnk} 
\usepackage{ulem}
\usepackage{amsmath}
\usepackage{amsfonts}
\usepackage[permil]{overpic}
\usepackage[export]{adjustbox}
\usepackage{pict2e} 
\usepackage{xcolor}

\title[Local Positional Encoding for Multi-Layer Perceptrons]%
      {Local Positional Encoding for Multi-Layer Perceptrons}

 \author[S. Fujieda \& A. Yoshimura \& T. Harada]
 {\parbox{\textwidth}{\centering S. Fujieda$^{1}$\orcid{0000-0002-2472-7365}
         and A. Yoshimura$^{1}$\orcid{0000-0001-5923-423X} 
         and T. Harada$^{1}$\orcid{0000-0001-5158-8455} 
         }
         \\
 {\parbox{\textwidth}{\centering $^1$Advanced Micro Devices, Inc.
        }
 }
 }

\begin{document}

\teaser{
    \centering
    \setlength{\tabcolsep}{0\linewidth}
    \renewcommand{\arraystretch}{1.0}
    \begin{tabular}{c c c@{\hspace{0.001\linewidth}} c c c@{\hspace{0.001\linewidth}} c c c@{\hspace{0.001\linewidth}} c c c}
        \multicolumn{3}{c}{ \begin{overpic}[width=0.245\linewidth]{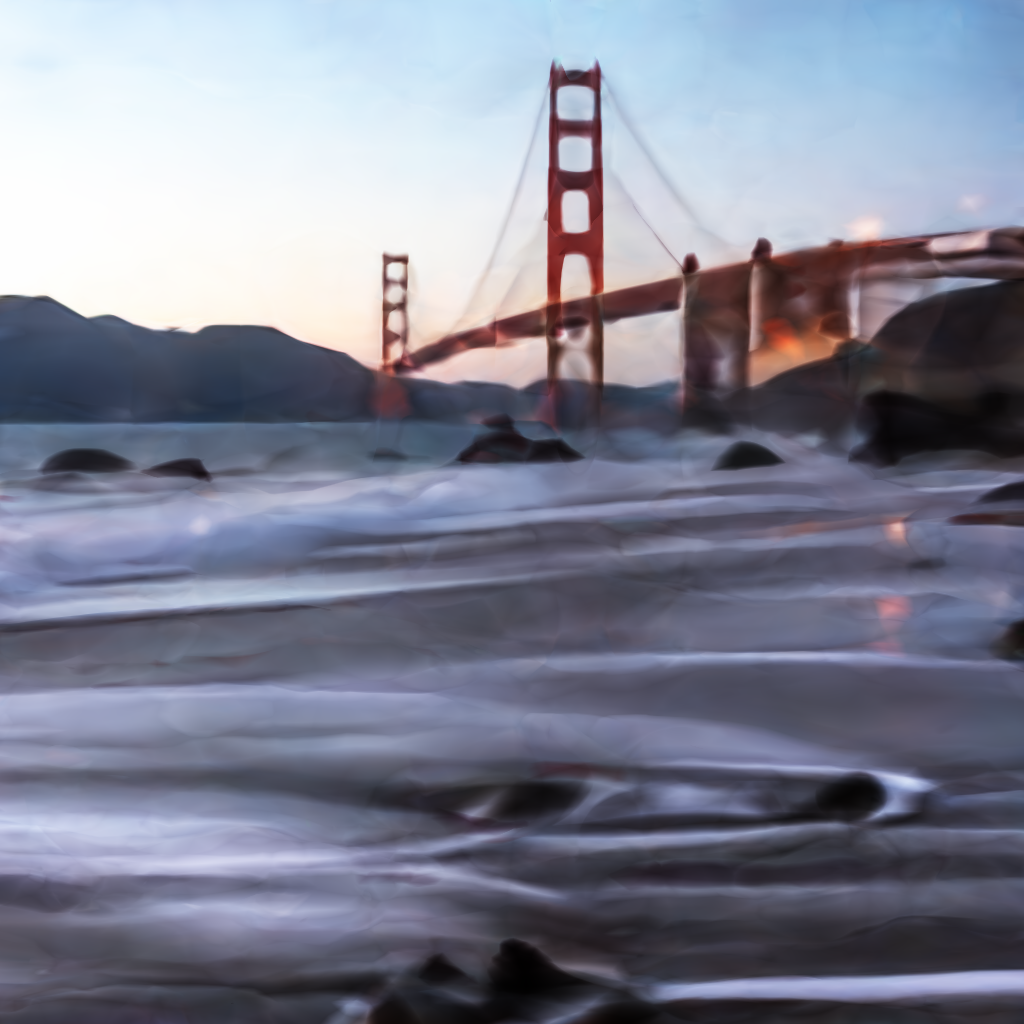} 
            \put(512,760){\linethickness{0.3mm}\color{red}\polygon(0,0)(118,0)(118,118)(0,118)}
            \put(666,594){\linethickness{0.3mm}\color{blue}\polygon(0,0)(118,0)(118,118)(0,118)}
            \put(810,657){\linethickness{0.3mm}\color{green}\polygon(0,0)(118,0)(118,118)(0,118)}
        \end{overpic} } &
        \multicolumn{3}{c}{ \begin{overpic}[width=0.245\linewidth]{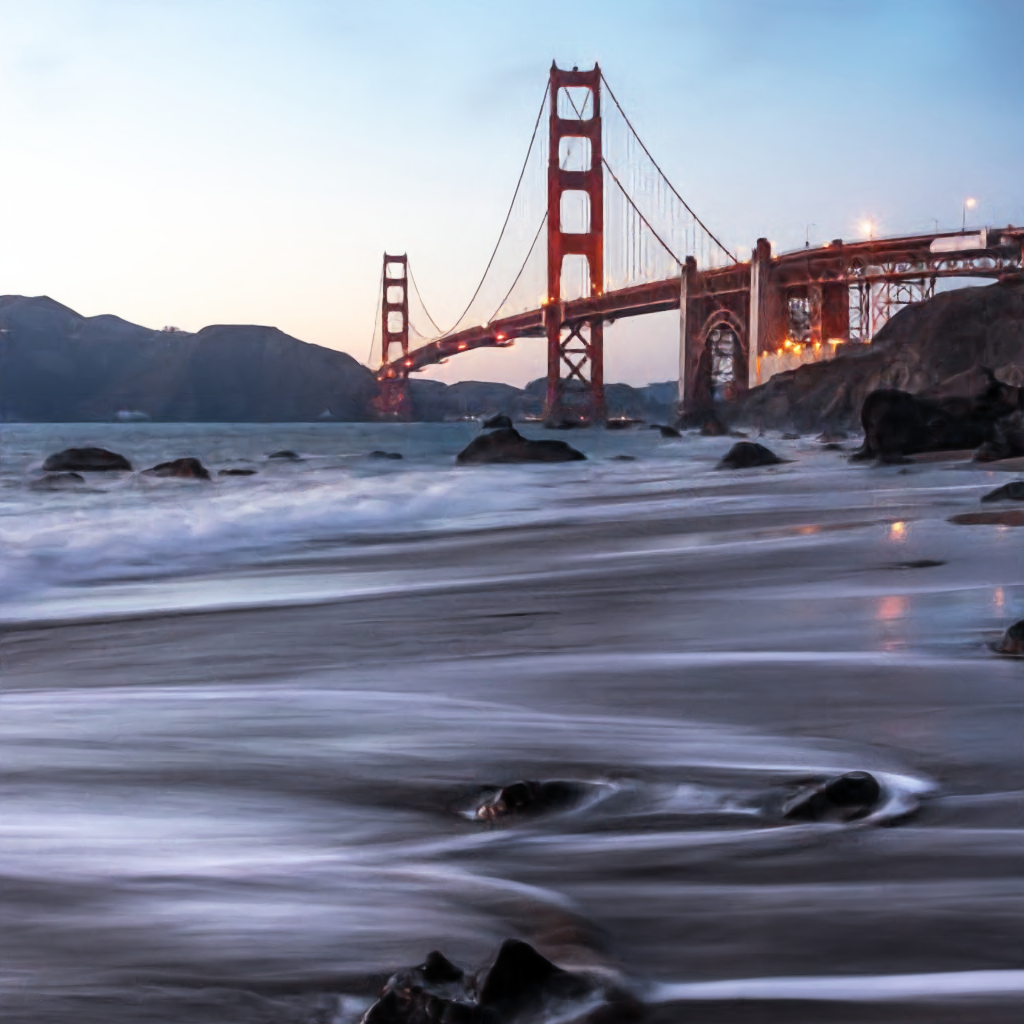} 
            \put(512,760){\linethickness{0.3mm}\color{red}\polygon(0,0)(118,0)(118,118)(0,118)}
            \put(666,594){\linethickness{0.3mm}\color{blue}\polygon(0,0)(118,0)(118,118)(0,118)}
            \put(810,657){\linethickness{0.3mm}\color{green}\polygon(0,0)(118,0)(118,118)(0,118)}
        \end{overpic} } &
        \multicolumn{3}{c}{ \begin{overpic}[width=0.245\linewidth]{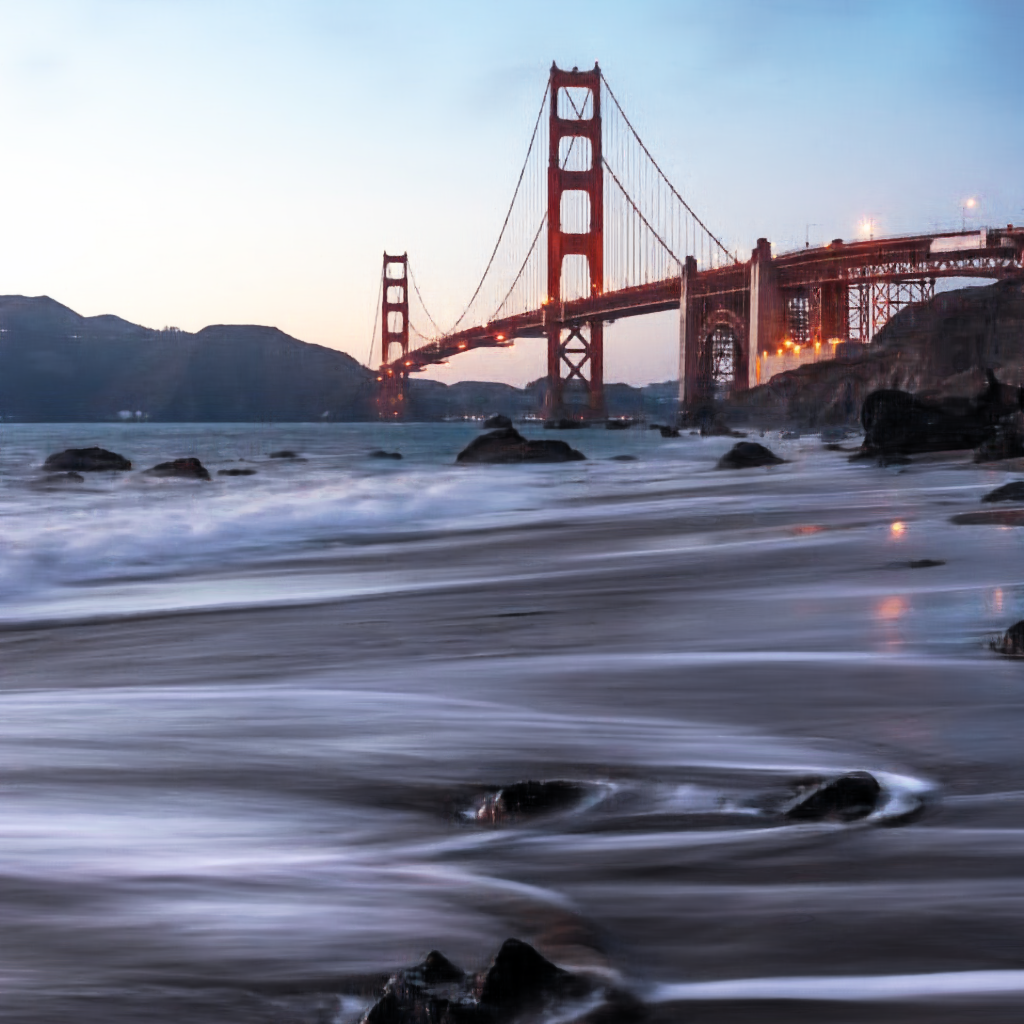} 
            \put(512,760){\linethickness{0.3mm}\color{red}\polygon(0,0)(118,0)(118,118)(0,118)}
            \put(666,594){\linethickness{0.3mm}\color{blue}\polygon(0,0)(118,0)(118,118)(0,118)}
            \put(810,657){\linethickness{0.3mm}\color{green}\polygon(0,0)(118,0)(118,118)(0,118)}
        \end{overpic} } &
        \multicolumn{3}{c}{ \begin{overpic}[width=0.245\linewidth]{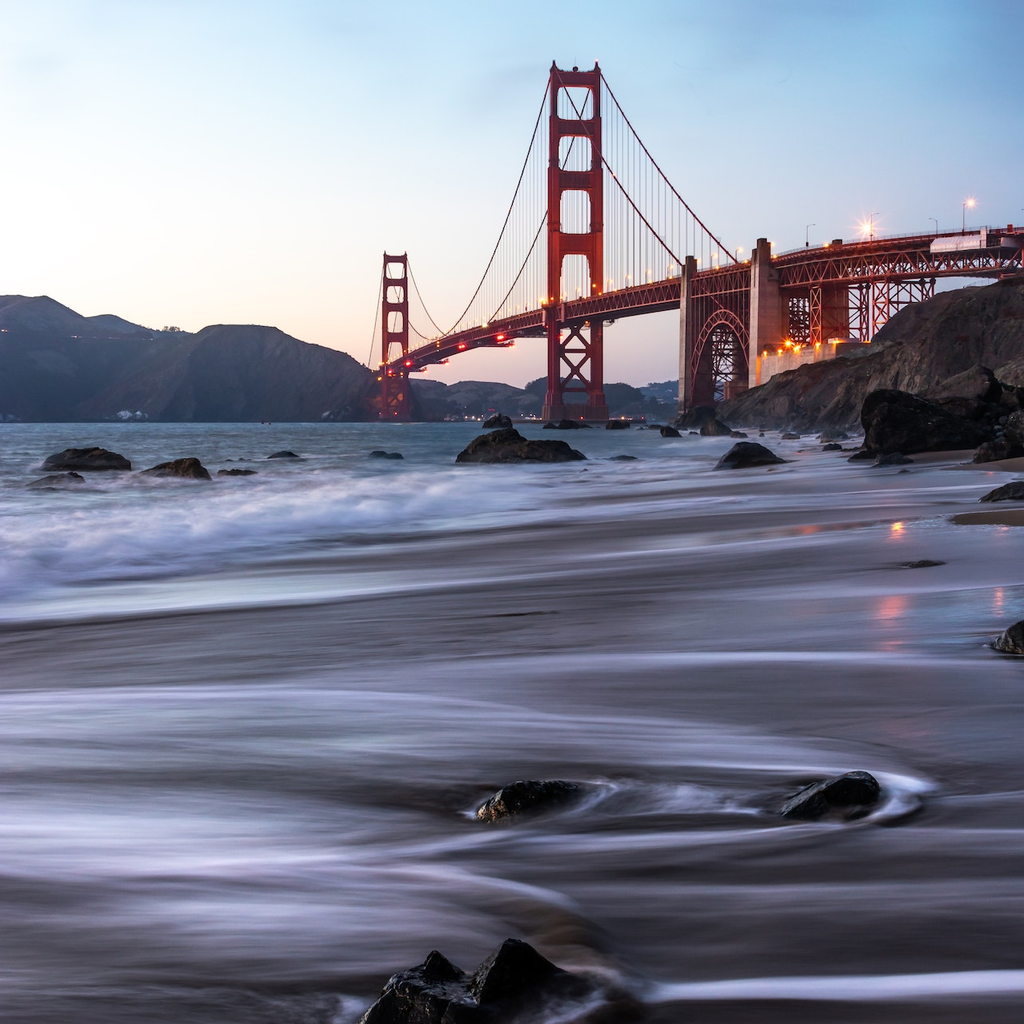} 
            \put(512,760){\linethickness{0.3mm}\color{red}\polygon(0,0)(118,0)(118,118)(0,118)}
            \put(666,594){\linethickness{0.3mm}\color{blue}\polygon(0,0)(118,0)(118,118)(0,118)}
            \put(810,657){\linethickness{0.3mm}\color{green}\polygon(0,0)(118,0)(118,118)(0,118)}
        \end{overpic} } \\

        \includegraphics[width=0.078\linewidth,viewport=524 778 642 896, clip, cfbox=red 1pt 0pt]{figs/teaser/bridge_pe.png} &
        \includegraphics[width=0.078\linewidth,viewport=682 608 800 726, clip, cfbox=blue 1pt 0pt]{figs/teaser/bridge_pe.png} &
        \includegraphics[width=0.078\linewidth,viewport=829 673 947 791, clip, cfbox=green 1pt 0pt]{figs/teaser/bridge_pe.png} &

        \includegraphics[width=0.078\linewidth,viewport=524 778 642 896, clip, cfbox=red 1pt 0pt]{figs/teaser/bridge_grid.png} &
        \includegraphics[width=0.078\linewidth,viewport=682 608 800 726, clip, cfbox=blue 1pt 0pt]{figs/teaser/bridge_grid.png} &
        \includegraphics[width=0.078\linewidth,viewport=829 673 947 791, clip, cfbox=green 1pt 0pt]{figs/teaser/bridge_grid.png} &

        \includegraphics[width=0.078\linewidth,viewport=524 778 642 896, clip, cfbox=red 1pt 0pt]{figs/teaser/bridge_lpe_noCellIdx.png} &
        \includegraphics[width=0.078\linewidth,viewport=682 608 800 726, clip, cfbox=blue 1pt 0pt]{figs/teaser/bridge_lpe_noCellIdx.png} &
        \includegraphics[width=0.078\linewidth,viewport=829 673 947 791, clip, cfbox=green 1pt 0pt]{figs/teaser/bridge_lpe_noCellIdx.png} &

        \includegraphics[width=0.078\linewidth,viewport=524 778 642 896, clip, cfbox=red 1pt 0pt]{figs/teaser/bridge_reference.png} &
        \includegraphics[width=0.078\linewidth,viewport=682 608 800 726, clip, cfbox=blue 1pt 0pt]{figs/teaser/bridge_reference.png} &
        \includegraphics[width=0.078\linewidth,viewport=829 673 947 791, clip, cfbox=green 1pt 0pt]{figs/teaser/bridge_reference.png} \\

        \multicolumn{3}{c}{ \small{PSNR: $25.58$ dB, SSIM: $0.7915$} } &
        \multicolumn{3}{c}{ \small{PSNR: $30.37$ dB, SSIM: $0.9133$} } &
        \multicolumn{3}{c}{ \small{PSNR: $31.40$ dB, SSIM: $0.9229$} } &
        \multicolumn{3}{c}{}  \\

        \multicolumn{3}{c}{ (a) Positional Encoding } & 
        \multicolumn{3}{c}{ (b) Grid Encoding } & 
        \multicolumn{3}{c}{ (c) Local Positional Encoding } &
        \multicolumn{3}{c}{ (d) Reference } \\
    \end{tabular}
    \caption{ Comparison of \textsc{Bridge} images reconstructed using an MLP with (a) positional encoding, (b) grid encoding, (c) local positional encoding, and (d) the reference. They all use a $64\times64$ grid with $16$-dimensional latent vectors and $4$ frequencies in positional encoding. The MLP has three hidden layers with $64$ neurons each. The original and reconstructed image resolutions are both $1,024\times1,024$. }
    \label{fig:teaser}
}

\maketitle
\begin{abstract}
A multi-layer perceptron (MLP) is a type of neural networks which has a long history of research and has been studied actively recently in computer vision and graphics fields. One of the well-known problems of an MLP is the capability of expressing high-frequency signals from low-dimensional inputs. There are several studies for input encodings to improve the reconstruction quality of an MLP by applying pre-processing against the input data. This paper proposes a novel input encoding method, local positional encoding, which is an extension of positional and grid encodings.
Our proposed method combines these two encoding techniques so that a small MLP learns high-frequency signals by using positional encoding with fewer frequencies under the lower resolution of the grid to consider the local position and scale in each grid cell.
We demonstrate the effectiveness of our proposed method by applying it to common 2D and 3D regression tasks where it shows higher-quality results compared to positional and grid encodings, and comparable results to hierarchical variants of grid encoding such as multi-resolution grid encoding with equivalent memory footprint.


\end{abstract}  
\section{Introduction} \label{Sec:intro}
A multi-layer perceptron (MLP) has been used in many applications in computer vision and graphics to find a mapping from a low-dimensional coordinate to other properties at that location.
However, MLPs usually suffer from capturing high-frequency signals from such low-dimensional inputs, which is known as \textit{spectral bias}~\cite{pmlr-v97-rahaman19a}.
One approach to handle this issue of an MLP is to map the input vector to a higher-dimensional space using positional encoding~\cite{tancik2020fourfeat, 10.1145/3503250}.
It applies sinusoidal functions to the input vector before passing it to the MLP.
Although positional encoding is a simple and effective approach, it requires a larger network as it increases the input dimension dramatically.
The highest reproducible frequency depends on the number of frequencies which is manually decided as a hyperparameter according to the use cases.
It has to be many enough to represent the desired high-frequency details.
To handle the number of frequencies efficiently, Hertz et al.~\cite{NEURIPS2021_4a06d868} introduce a novel learning policy, SAPE, to select proper frequencies according to the local spatial position to better fit the locally varying signals.
Mip-NeRF~\cite{9710056} and its extension, Mip-NeRF 360~\cite{9878829}, also try to tune the number of frequencies automatically by using features approximating the integral over the positional encoding of all coordinates within a sub-volume.
If a particular frequency has a period which is larger than the size of the sub-volume, they penalize the encoding of that frequency as its amplitude gets close to zero.
These methods have a better capability of representing high-frequency signals but still require high-dimensional inputs which lead to a large MLP.

The other approach to resolve the same problem of an MLP is to apply grid encoding to the inputs~\cite{sitzmann2019deepvoxels, chibane20ifnet, 10.1145/3478513.3480569, 10.1145/3450626.3459795, mehta2021modulated, martel2021acorn, 10.1145/3528223.3530127, 10.1145/3528233.3530727, Weier:2023}.
The core idea of grid encoding is to prepare one or multiple grids overlaying the input domain and store latent vectors in each grid cell.
We can classify this as another input encoding method, but how it uses an MLP is different from positional encoding.
Grid encoding learns features on grid cells and uses an MLP as a decoder, which makes its network smaller.
This is a reason for the faster training compared to positional encoding.
Another nature of grid encoding is that it trades off the network complexity for the storage space.
Therefore, it requires a higher-resolution grid or a higher-dimensional latent vector for each cell to resolve higher-frequency signals, which need larger memory space.
Higher-dimensional inputs also require a higher-dimensional grid.
Even with three-dimensional inputs such as a position in the 3D space, the memory overhead is quite significant.
This nature makes it memory-intensive to use grid encoding for a problem with high-frequency information in the high-dimensional domain.
There are some studies that try to overcome this by using a sparse voxel octree holding features~\cite{Takikawa2021NeuralGL}, introducing a fixed-size hash table in a pyramid of grids~\cite{10.1145/3528223.3530127}, or storing indices into the feature codebook in each grid cell~\cite{10.1145/3528233.3530727}.
They lower the memory pressure but do not solve the fundamental problem of grid encoding which requires higher-resolution grids to achieve good quality.

Additionally, Zip-NeRF~\cite{barron2023zipnerf} tries to combine these two input encodings to take advantage of both benefits: positional encoding offers the simple representation of high-frequency signals and grid encoding allows us to use a small MLP.
It integrates a pyramid of grids with hash tables~\cite{10.1145/3528223.3530127} into Mip-NeRF 360's framework~\cite{9878829}.
However, it is still memory-intensive to adopt a set of grids and requires a large MLP.
Also, Karnewar et al.~\cite{ReluField_sigg_22} explore the potential of grid encoding from a different perspective.
They introduce ReLU Fields where a simple ReLU function is applied to the interpolated grid values without any neural network as a decoder, which results in faster training and evaluation.
However, it cannot model more than one discontinuity on signals per grid cell, so other signals such as natural images cannot be represented.

There is a concurrent work by Vaidyanathan et al. ~\cite{ntc2023} which addresses the same problem by a similar approach to our work released in the same year to this work. The difference is how to use the grid values. Our method uses the grid values as amplitudes of sinusoidal encodings by multiplying them together while they concatenate them.

In this work, we propose local positional encoding for an MLP which is a hybrid of positional encoding and grid encoding.
Local positional encoding can resolve high-frequency signals without using as many frequencies as positional encoding requires, and without preparing a high-resolution grid as grid encoding requires.
Our method also stores latent coefficients in each grid cell which are combined with the signal of the local coordinate in the cell mapped by positional encoding with a few frequencies.
Therefore, considering its local position and scale in each grid cell, our proposed method can resolve higher-frequency information than positional encoding with the same number of frequencies and grid encoding with the same resolution size as shown in Fig.~\ref{fig:teaser}.
We evaluate local positional encoding in two applications, 2D image reconstruction and 3D signed distance functions, to present the advantage of the proposed method.

\begin{figure}
    \centering
    \includegraphics[width=0.95\columnwidth]{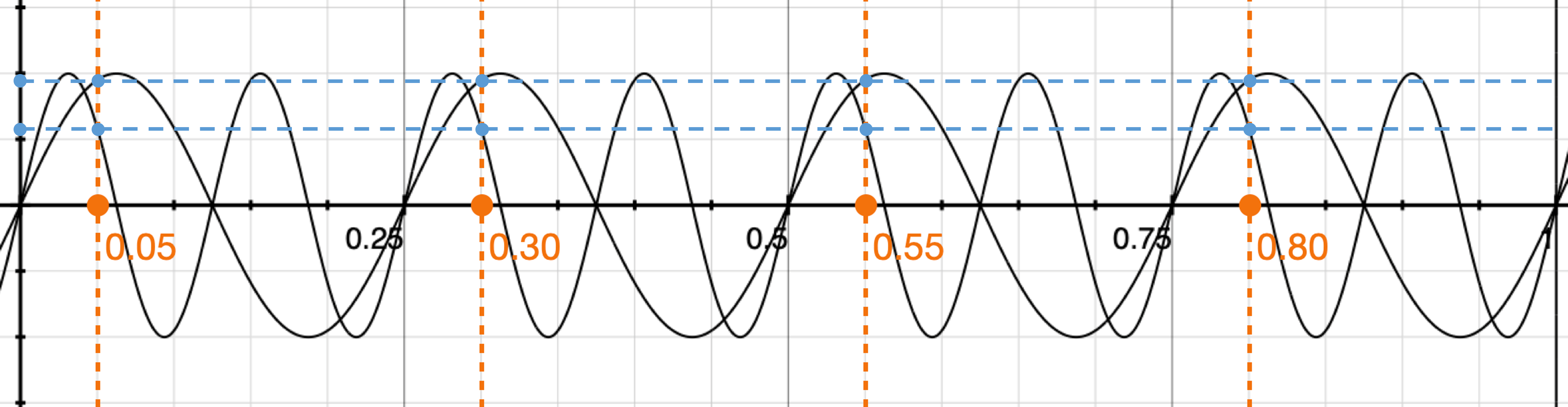}
    \caption{Problem of using only higher frequencies in positional encoding. There are multiple points mapped to the same vector. In this one-dimensional example, $x=0.05, 0.30, 0.55, 0.80$ are mapped to exactly the same vector from which a neural network cannot handle the difference.}
    \label{fig:aliasing}
\end{figure}
\section{Method}
\subsection{Positional Encoding} \label{Sec:freqEncoding}
Positional encoding transforms the input vector $x$ to a higher-dimensional vector by the following equation:
\begin{multline}
\label{eq:FE}
PE(x) = [\cos(2^0\pi x), \sin(2^0\pi x), \cdots, \\
        \cos(2^{n-1}\pi x),  \sin(2^{n-1}\pi x)],
\end{multline}
where $n$ is the number of frequencies used to encode the signal. 
If we use positional encoding with a limited number of frequencies (e.g. 4 frequencies), it simply fails to encode high-frequency signals as shown in Fig.~\ref{fig:teaser}a.
Larger $n$ could be able to capture higher-frequency details of the input signal but increases the input dimension of an MLP.
An offset to the frequencies (i.e. $\sin(2^{o+i}\pi x)$ where $o$ is the offset which we usually start from $0$) can decrease the input dimension; however, this approach lets multiple locations in the domain mapped to the same vector, as illustrated in Fig.~\ref{fig:aliasing}.
It leads to training failures since the network cannot distinguish such inputs.
In other words, the uniqueness of each feature vector is lost with an offset to the frequencies. 
Another approach to capturing high-frequency information with fewer frequencies would be to subdivide the domain of the signal into cells and assign a network with positional encoding to each cell.
However, in this case, the total memory consumption increases proportionally to the number of cells.
Instead, to reduce the size of the entire network, we combine the ideas of positional and grid encodings to provide hints to the network to identify the cells. 

\subsection{Local Positional Encoding} \label{Sec:lpe}
Our proposed method, local positional encoding, uses a uniform grid to define cells as other grid encoding methods.
In order to capture high-frequency information with low memory consumption, we use a single global network for all the cells and introduce a weight for each element of positional encoding stored in each cell to utilize the local information in the cell.
We call the weights latent coefficients and train them along with other network weights.
Our approach is also inspired by the Short-time Fourier transform (STFT) which provides the frequency information localized in a short term when the frequency of a signal varies over time by applying a window function which spans only for a short period of time to the Fourier transform.
In our method, the latent coefficients control the amplitudes of sinusoidal encodings in each cell so that the local frequency information in the cell is captured for the spatially varying signals.

More specifically, local positional encoding starts with a transformation of the global coordinate of the input point to the local coordinate:
\begin{align}
    x_l & = x_g \cdot N - z, \\
    z   & = \lfloor x_g \cdot N \rfloor,
\end{align}
where $x_l, x_g \in \mathbb{R}^d$ are local and global coordinates, $N$ is the grid resolution and $z$ is the cell index.
The cell index $z$ is also used to look up the latent coefficients for the cell:
\begin{equation}
    A_{PE}(z) = [ a^c_0(z), a^s_0(z), \cdots, a^c_{n-1}(z), a^s_{n-1}(z) ],
\end{equation}
where the subscripts $c$ and $s$ denote the coefficients for cosine and sine functions, respectively.
Then we apply positional encoding to the local coordinate $x_l$ to generate a feature vector which is multiplied by the latent coefficients $A_{PE}(z)$.
We can see that this operation limits the effect of the sinusoidal functions for each cell to some range where trainable latent coefficients work as a window function for STFT.
Thus, the input to the MLP is computed with the element-wise multiplication of $PE(x_l)$ and $A_{PE}(z)$:
\begin{multline}
    \label{eq:LPE1}
    PE(x_l) \odot A_{PE}(z) = [a^c_0(z) \cos(2^0\pi x_l), a^s_0(z) \sin(2^0\pi x_l), \cdots, \\
                                a^c_{n-1}(z) \cos(2^{n-1}\pi x_l), a^s_{n-1}(z) \sin(2^{n-1}\pi x_l)].
\end{multline}

However, using this feature vector computed with Equation~\ref{eq:LPE1} causes visual discontinuities at the cell boundaries.
This is because the cosine function with the lowest frequency (i.e. $\cos{(2^0 \pi x)}$) has discontinuities at the edges of the range $[0, 1]$.
Therefore, instead of using the sinusoidal encodings with the lowest frequency, $a^c_0(z) \cos{(2^0 \pi x_l)}$ and $a^s_0(z) \sin{(2^0 \pi x_l)}$, we use the two-dimensional latent coefficient $A_G(z) = [a^g_0(z), a^g_1(z)]$ stored in the grid cell for the input of the MLP.
$A_G(z)$ works as the feature vector of grid encoding without being multiplied by sinusoidal functions.
As a result, local positional encoding transforms the input vector $x_g$ as the following equation:
\begin{equation}
\label{eq:LPE2}
\begin{split}
    LPE(x_g) = & [A_G(z), a^c_1(z) \cos(2^1\pi x_l), a^s_1(z) \sin(2^1\pi x_l), \cdots, \\
               & a^c_{n-1}(z) \cos(2^{n-1}\pi x_l), a^s_{n-1}(z) \sin(2^{n-1}\pi x_l)].
\end{split}
\end{equation}
And the trainable latent coefficients stored in each cell are: 
\begin{equation}
    \label{eq:latentCoeff}
    A(z) = [ A_G(z), a^c_1(z), a^s_1(z), \cdots, a^c_{n-1}(z), a^s_{n-1}(z) ].
\end{equation}

\begin{figure}
    \centering
    \includegraphics[width=\columnwidth]{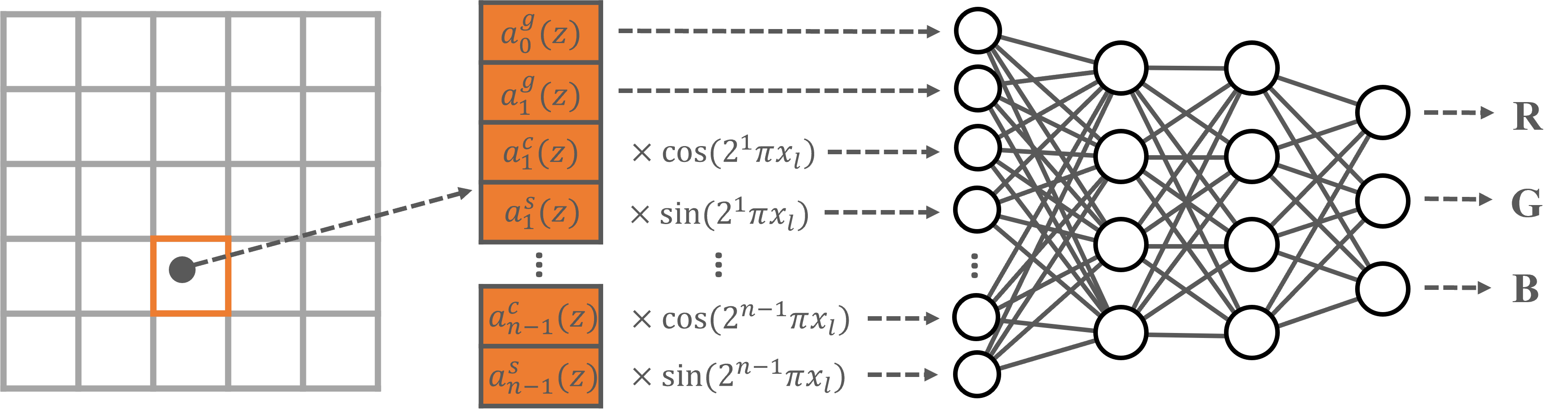}
    \caption{Illustration of local positional encoding in a two-dimensional image reconstruction problem.}
    \label{fig:pipeline}
\end{figure}

Additionally, assigning one set of the latent coefficients in a cell causes discontinuities at the edges of the cell.
Thus, we store the latent coefficients at each corner vertex of the cell and they are linearly interpolated based on the local coordinate $ x_l $ to avoid discontinuities.
Fig.~\ref{fig:pipeline} illustrates the example of local positional encoding in a 2D image reconstruction problem.

\subsection{Network}
Our network is a small MLP with three hidden layers with $64$ neurons each.
We can use any activation in the MLP, but we experimentally choose a leaky ReLU activation function with $\alpha = 0.01$ in all the examples in this paper, except for the output layer to which we apply a sigmoid activation function for the image reconstruction and none for signed distance functions. 
The dimension of the input layer is given by $2 \cdot n \cdot d$ where $d$ is the dimension of the input vector. 
When we use $4$ frequencies in local positional encoding, for example, each cell in the grid stores an $8$-dimensional feature vector for each dimension, following Equation~\ref{eq:latentCoeff}.
Therefore, the input to the network is also an $8$-dimensional vector computed with Equation~\ref{eq:LPE2} in a one-dimensional problem.

The weights of the neural network are initialized with Xavier initialization procedure~\cite{Glorot2010UnderstandingTD}.
The latent coefficients in grid cells are initialized with the uniform distribution $\mathcal{U}(-10^{-4}, 10^{-4})$.

\begin{table*}
    \caption{Comparisons of PSNR and SSIM for the images shown in Fig.~\ref{fig:teaser} and Fig.~\ref{fig:table}. "PE" is short for "Positional Encoding", "GE" is for "Grid Encoding" and "LPE" is for "Local Positional Encoding". The bold numbers show the best results for each image. }
    \label{tab:image_quantitative}
    \centering
    \small
    \begin{tabular}{p{15mm}||c c c|c c c|c c c|c c c}
    \toprule 
        & \multicolumn{3}{c|}{\textsc{Bridge}}  & \multicolumn{3}{c|}{\textsc{Keyboard}}  & \multicolumn{3}{c|}{\textsc{Fish Market}} & \multicolumn{3}{c}{\textsc{Trees}} \\
    \midrule
        & PE & GE & LPE & PE & GE & LPE & PE & GE & LPE & PE & GE & LPE \\ 
    \midrule
    PSNR [dB] & 25.58  & 30.37  & \textbf{31.40}  & 27.98  & 35.26  & \textbf{35.98}  & 18.80  & 23.12  & \textbf{24.03}  & 20.56  & 24.20  & \textbf{24.54}  \\
    SSIM      & 0.7915 & 0.9133 & \textbf{0.9229} & 0.8375 & 0.9359 & \textbf{0.9367} & 0.4608 & 0.7065 & \textbf{0.7071} & 0.4629 & 0.7090 & \textbf{0.7167} \\
    \bottomrule
    \end{tabular}
\end{table*}

\subsection{Optimization}
Local positional encoding does not add any trainable parameter to the network.
However, we introduce the latent coefficients, $ A(z) $, in the grid which are trainable parameters to be updated during the training using the following equations:
\begin{align}
\frac{\partial \mathcal{L}}{\partial a^c_i} & = \frac{\partial \mathcal{L}}{\partial n_i} \cdot \frac{\partial n_i}{\partial a^c_i} = \cos(2^i \pi x_l) \cdot \frac{\partial \mathcal{L}}{\partial n_i}, \\
\frac{\partial \mathcal{L}}{\partial a^s_i} & = \frac{\partial \mathcal{L}}{\partial n_i} \cdot \frac{\partial n_i}{\partial a^s_i} = \sin(2^i \pi x_l) \cdot \frac{\partial \mathcal{L}}{\partial n_i},
\end{align}
where $\mathcal{L}$ is a reconstruction loss and $n_i$ is the corresponding sinusoidal encoding of the input vector to the MLP in Equation~\ref{eq:LPE2}.
We jointly optimize the network and the grid using gradient descent with the Adam optimizer~\cite{Adam}, where we set $\beta_1 = 0.9$, $\beta_2 = 0.999$ and $\epsilon = 10^{-15}$.
For the image reconstruction task, we use the $\mathcal{L}^2$ loss function with a learning rate of $0.02$.
And for the signed distance functions, we use the mean absolute percentage error (MAPE) similar to~\cite{10.1145/3528223.3530127} with a learning rate of $10^{-4}$.
These hyperparameters are chosen experimentally.

\section{Results}
\label{sec:results}

We implemented an MLP with different encodings using C++ and HIP for GPU programming language \cite{hip}. All the evaluations are done by executing the codes on AMD Radeon\textsuperscript{\texttrademark} RX 7900 XT or AMD Radeon\textsuperscript{\texttrademark} RX 7900 XTX GPU on Windows machines.
All the training weights are stored in 32-bit floating point values.
We evaluate local positional encoding against other input encodings, such as positional encoding and grid encoding, in the image reconstruction task (Sec.~\ref{sec:image}) and in the task of representing signed distance functions (Sec.~\ref{sec:sdf}).
Also, we compare our encoding with multi-resolution grids for signed distance functions (Sec.~\ref{sec:cmpMulti}).

\subsection{Image Reconstruction} \label{sec:image}

\begin{figure}
    \centering
    \setlength{\tabcolsep}{0.002\linewidth}
    \begin{tabular}{ccccc}
        \includegraphics[width=0.4\linewidth]{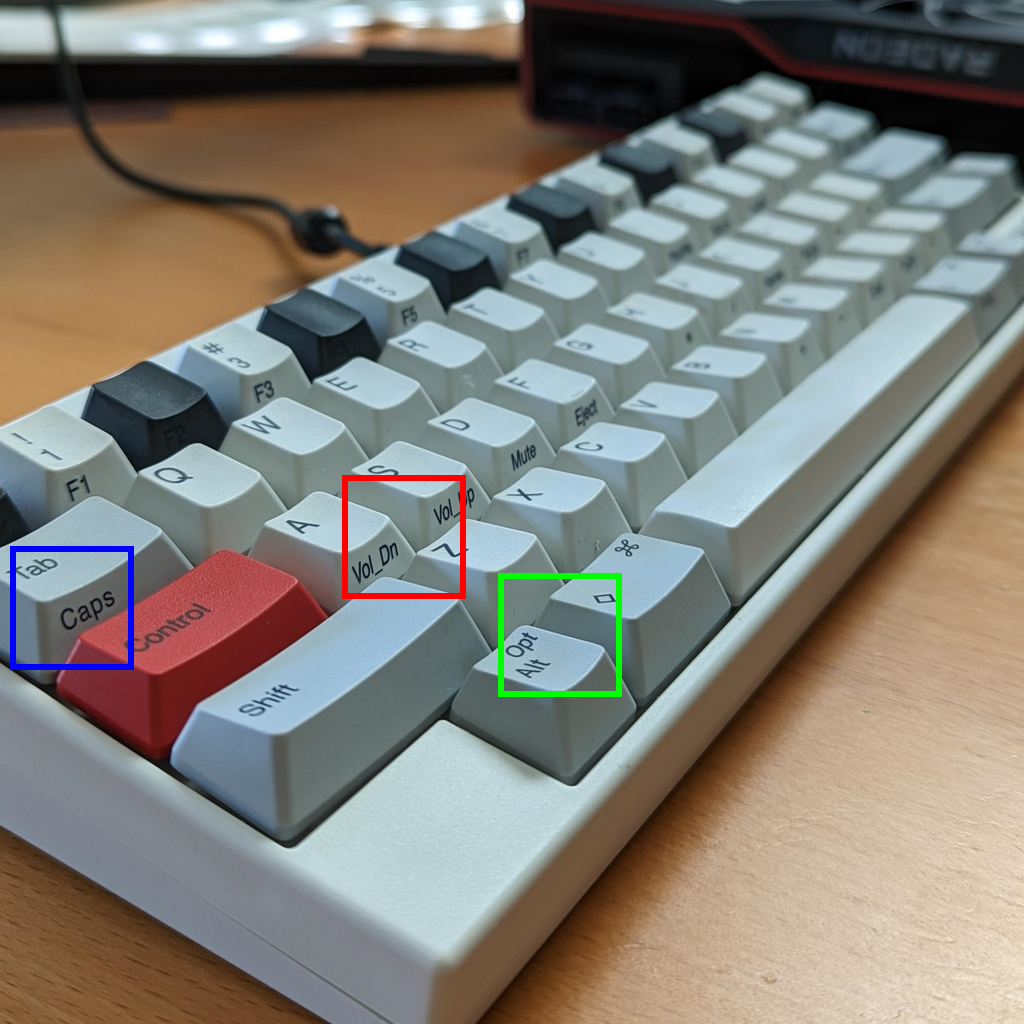} &
        \includegraphics[width=0.135\linewidth]{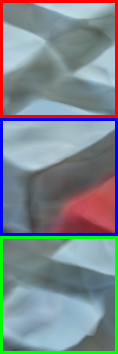} &
        \includegraphics[width=0.135\linewidth]{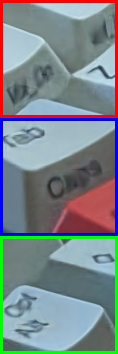} &
        \includegraphics[width=0.135\linewidth]{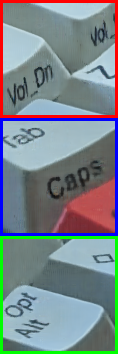} &
        \includegraphics[width=0.135\linewidth]{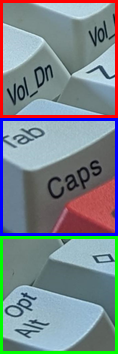} \\
        
        & \small{(a) PE} & \small{(b) GE} & \small{(c) LPE} & \small{(d) Ref.}
    \end{tabular}
    \caption{Detailed comparison of \textsc{Keyboard} images. (a) Positional encoding, (b) grid encoding, (c) local positional encoding, and (d) the reference.}
    \label{fig:zoomup}
\end{figure}

\begin{figure*}
    \centering
    \setlength{\tabcolsep}{0.002\linewidth}
    \begin{tabular}{ccccc}
        \raisebox{0.12\linewidth}{\rotatebox[origin = c]{90}{\textsc{Keyboard}}} &
        \includegraphics[width=0.24\linewidth]{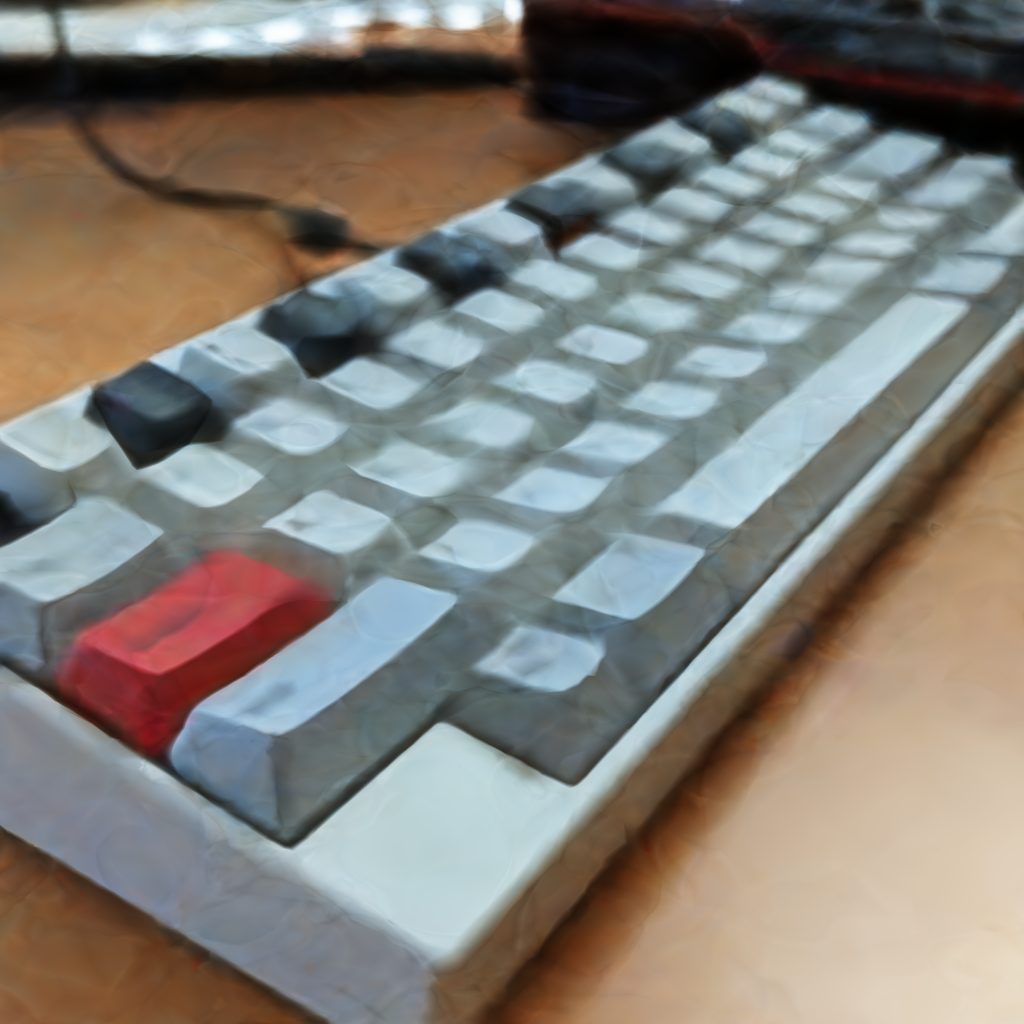} &
        \includegraphics[width=0.24\linewidth]{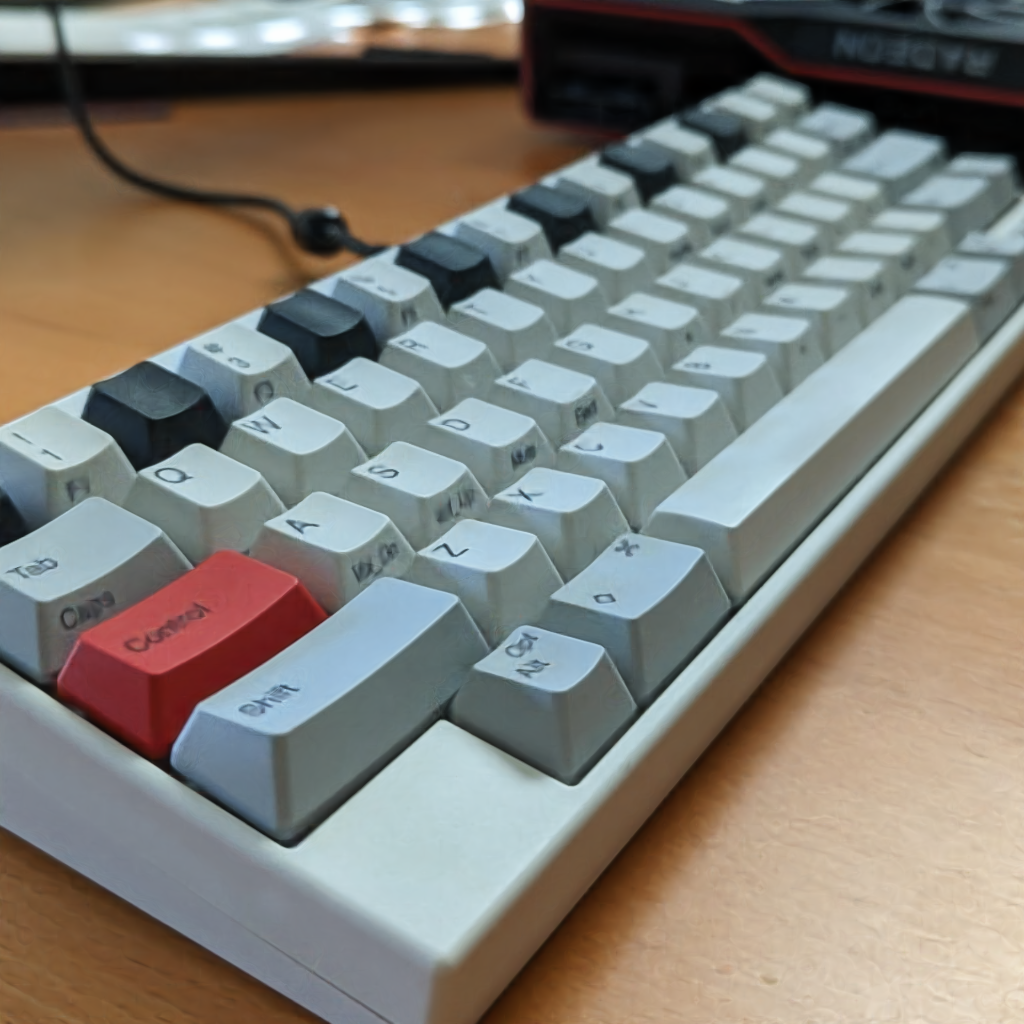} &
        \includegraphics[width=0.24\linewidth]{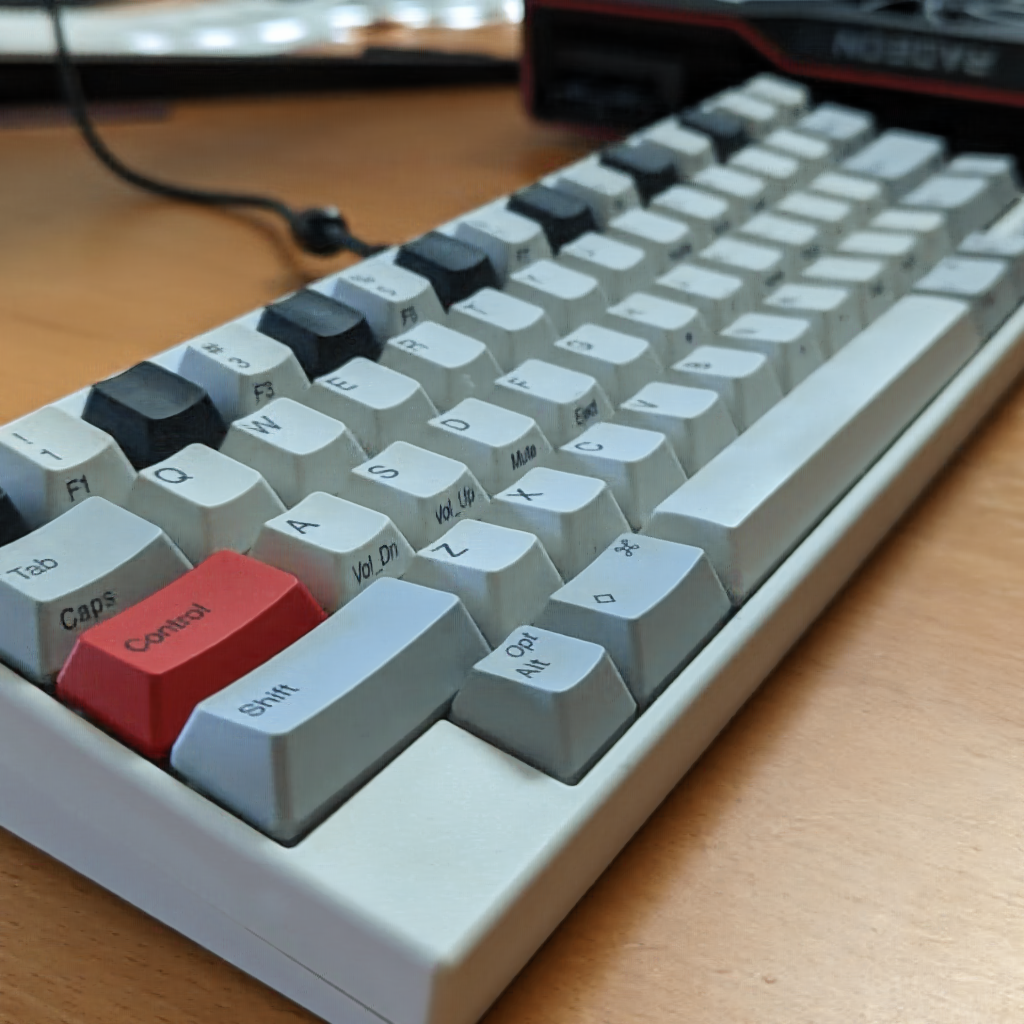} &
        \includegraphics[width=0.24\linewidth]{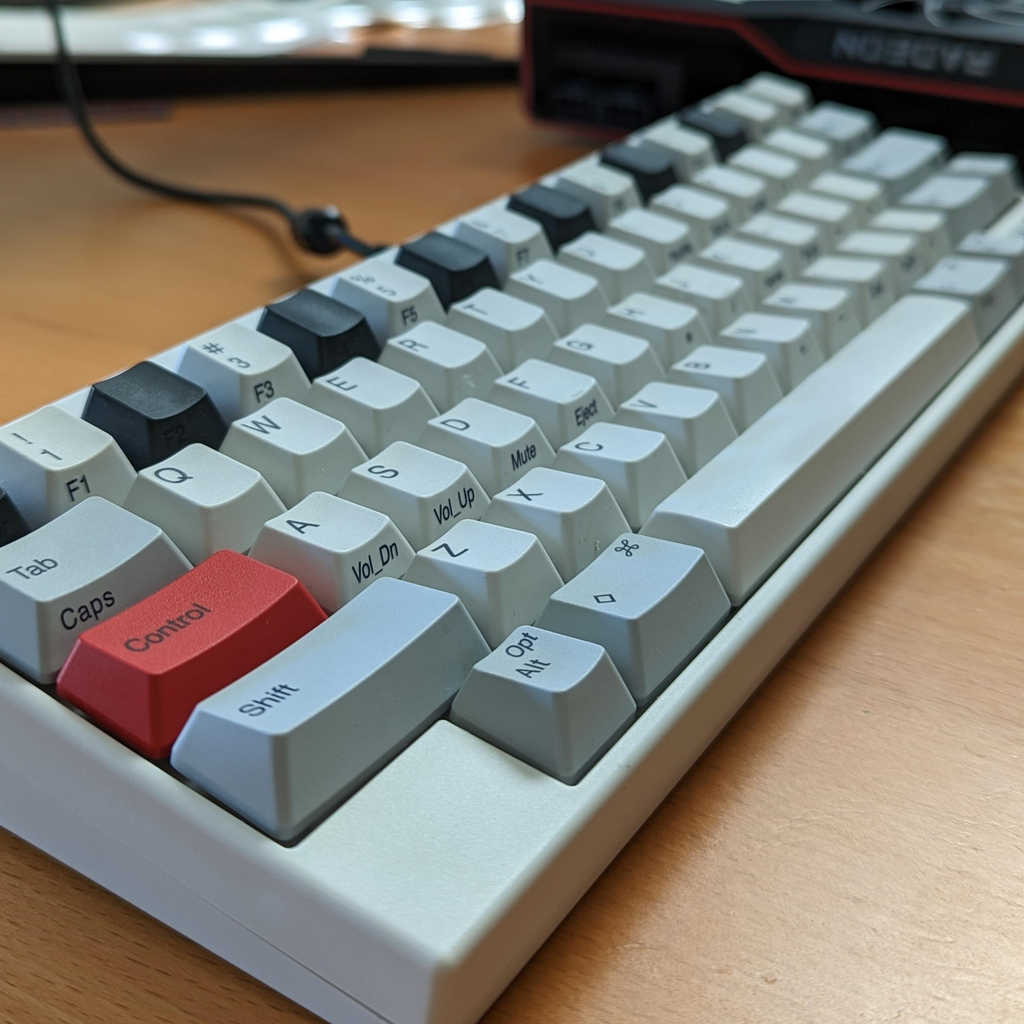} \\

        \raisebox{0.12\linewidth}{\rotatebox[origin = c]{90}{\textsc{Fish Market}}} &
        \includegraphics[width=0.24\linewidth]{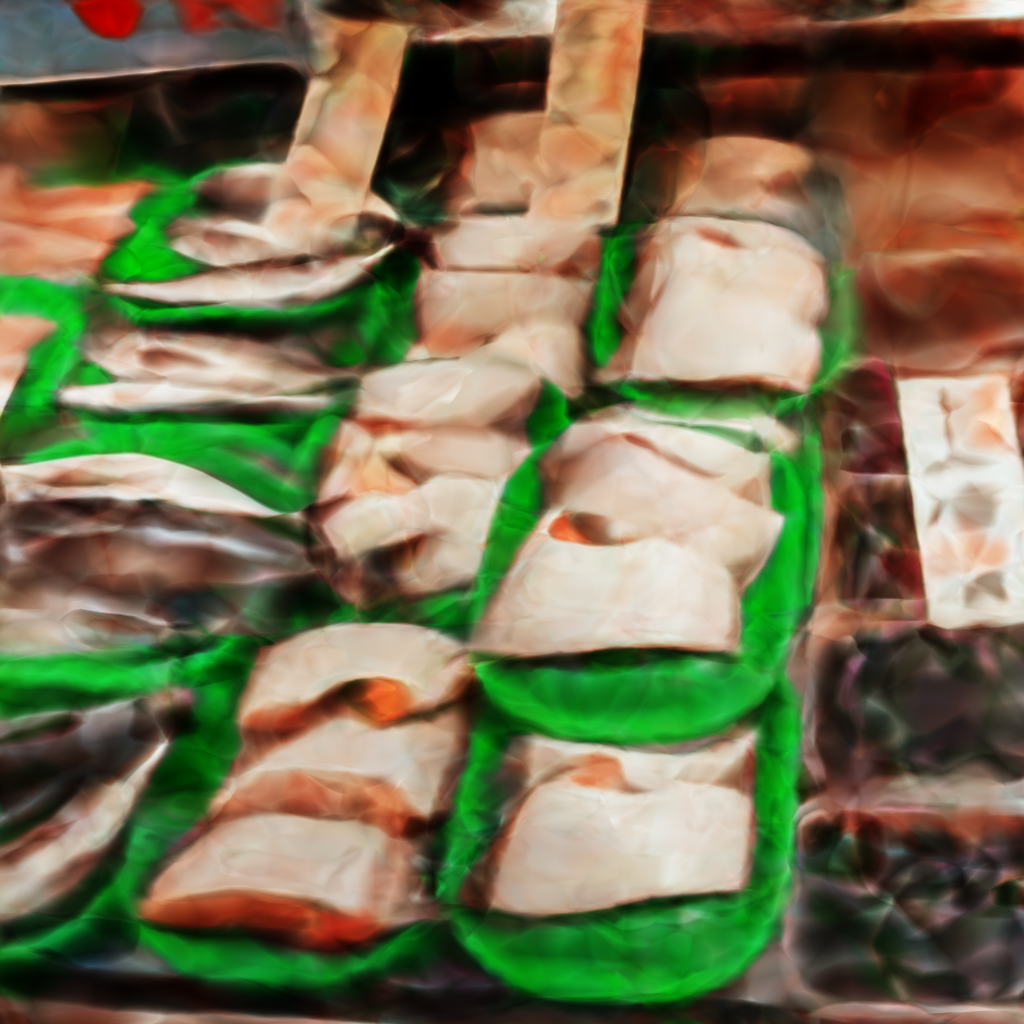} &
        \includegraphics[width=0.24\linewidth]{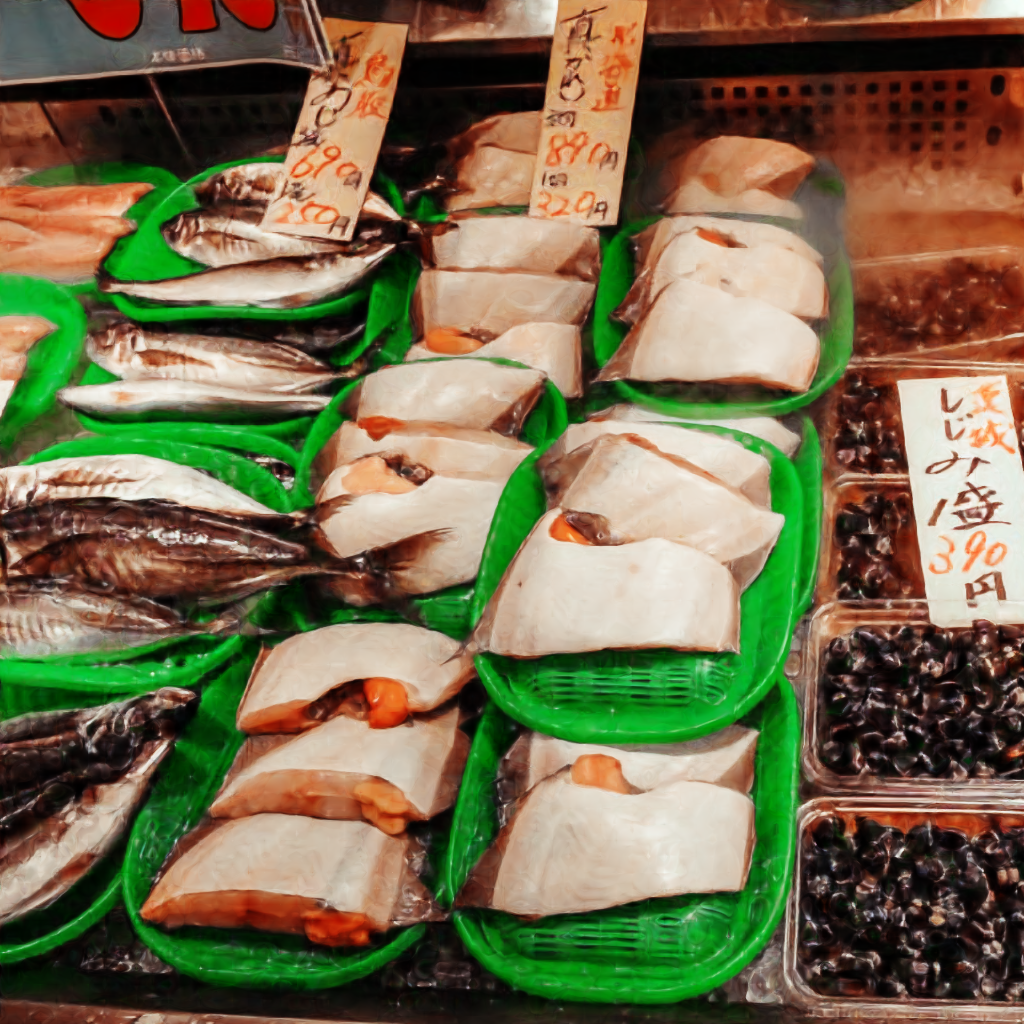} &
        \includegraphics[width=0.24\linewidth]{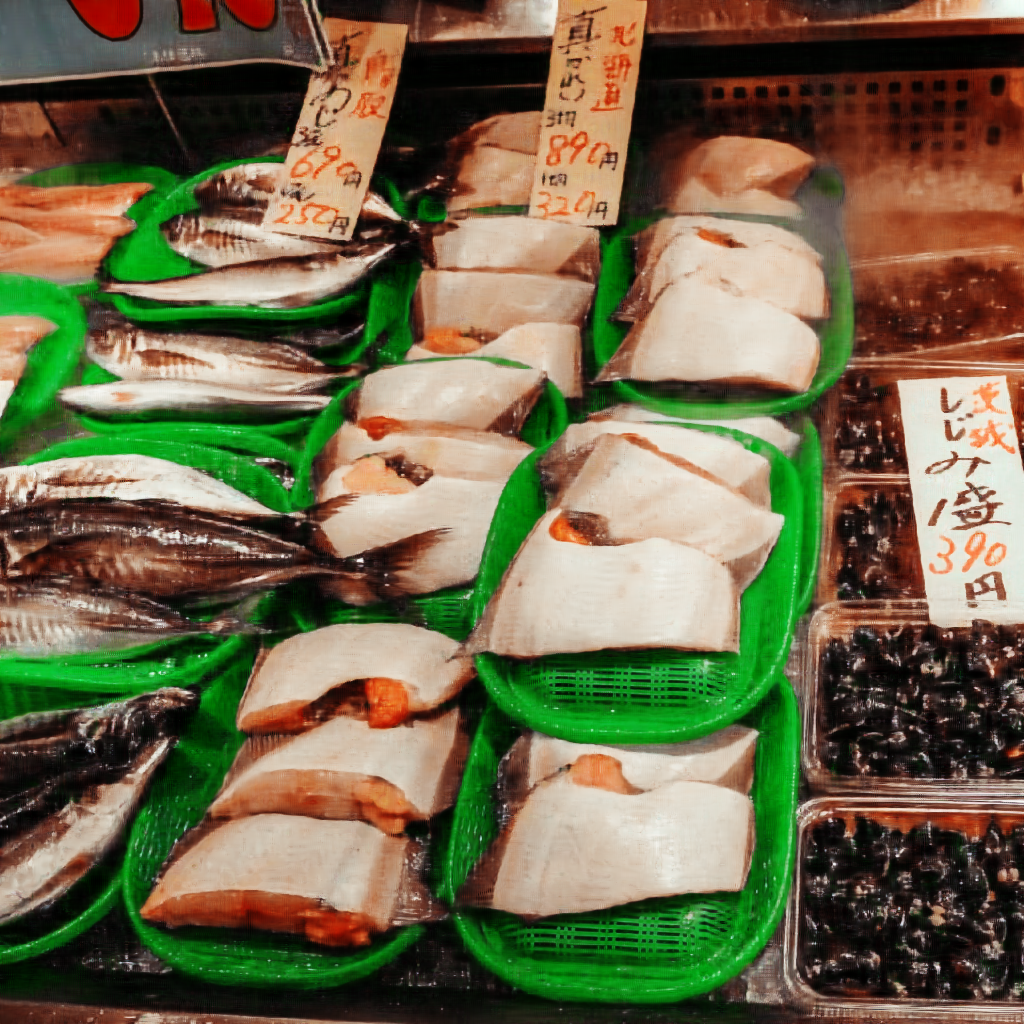} &
        \includegraphics[width=0.24\linewidth]{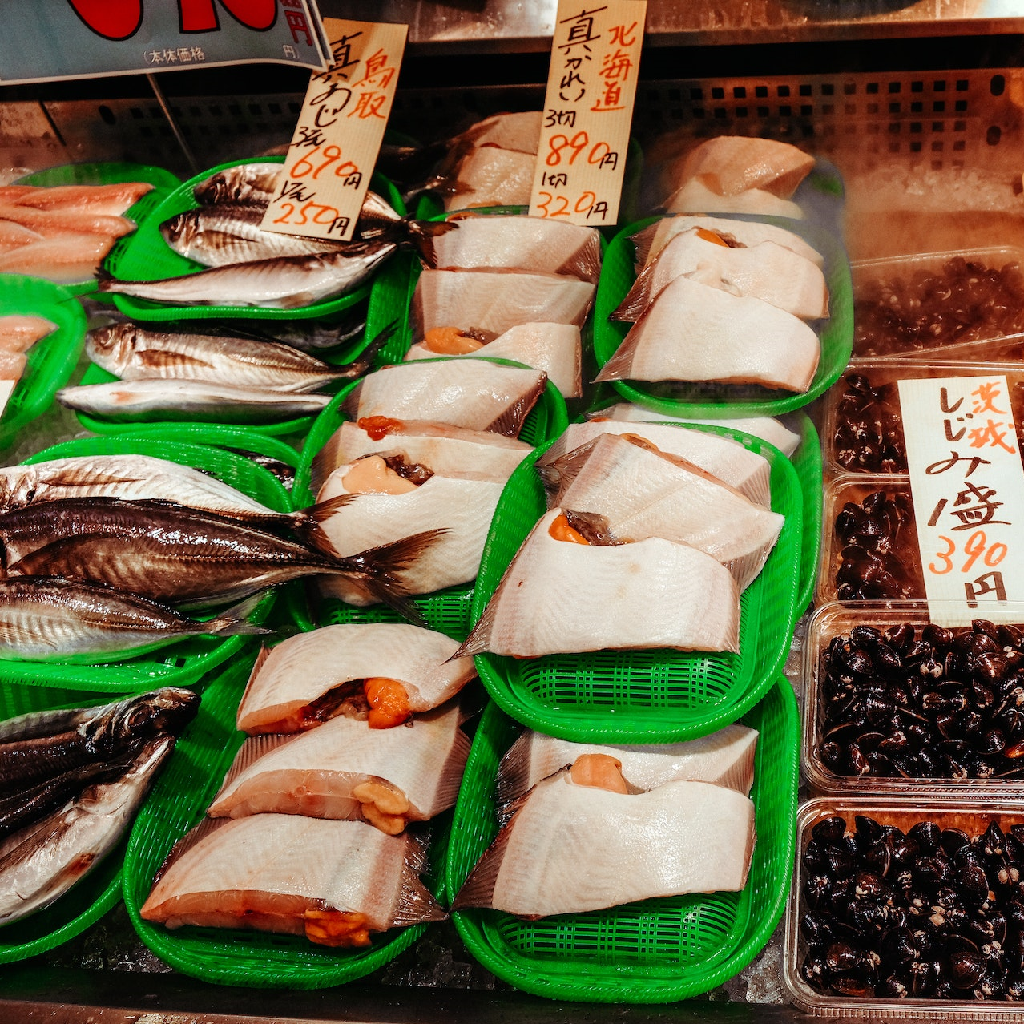} \\

        \raisebox{0.11\linewidth}{\rotatebox[origin = c]{90}{\textsc{Trees}}} &
        \includegraphics[width=0.24\linewidth]{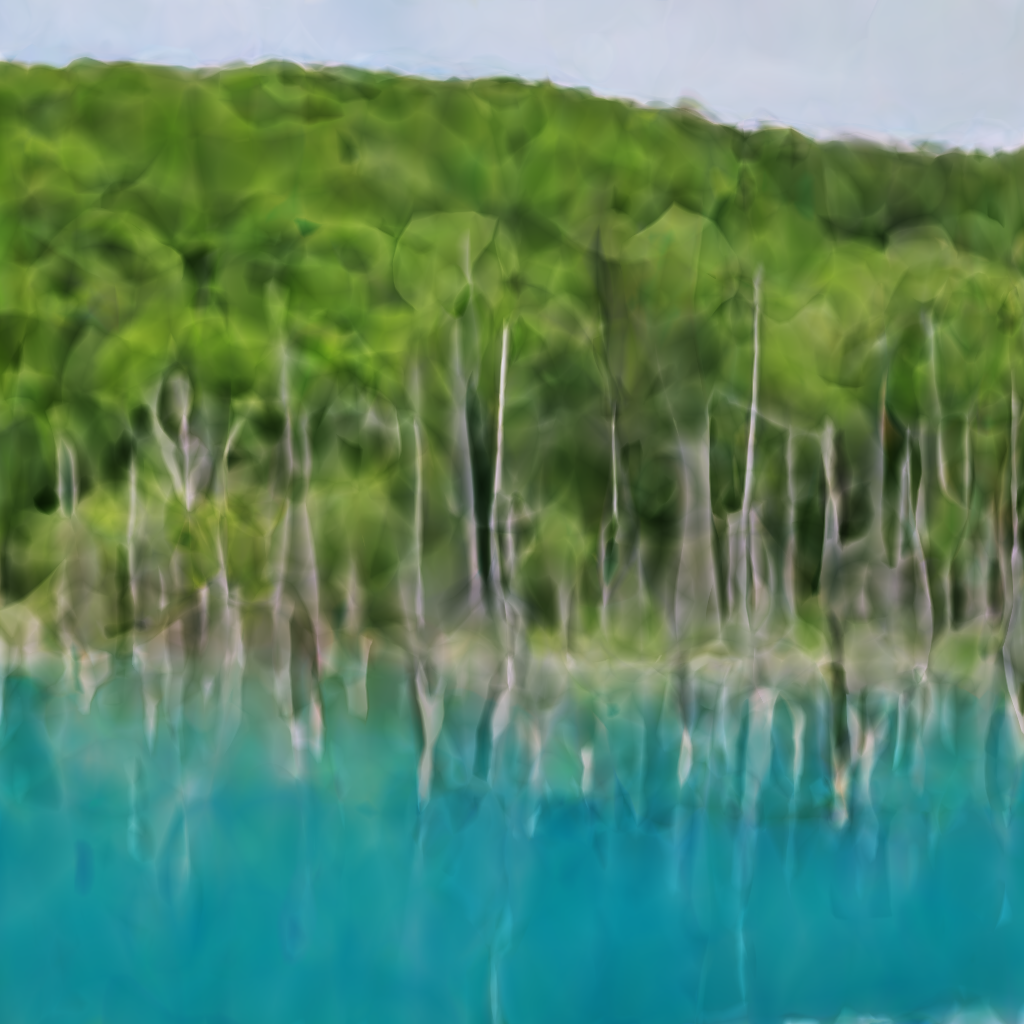} &
        \includegraphics[width=0.24\linewidth]{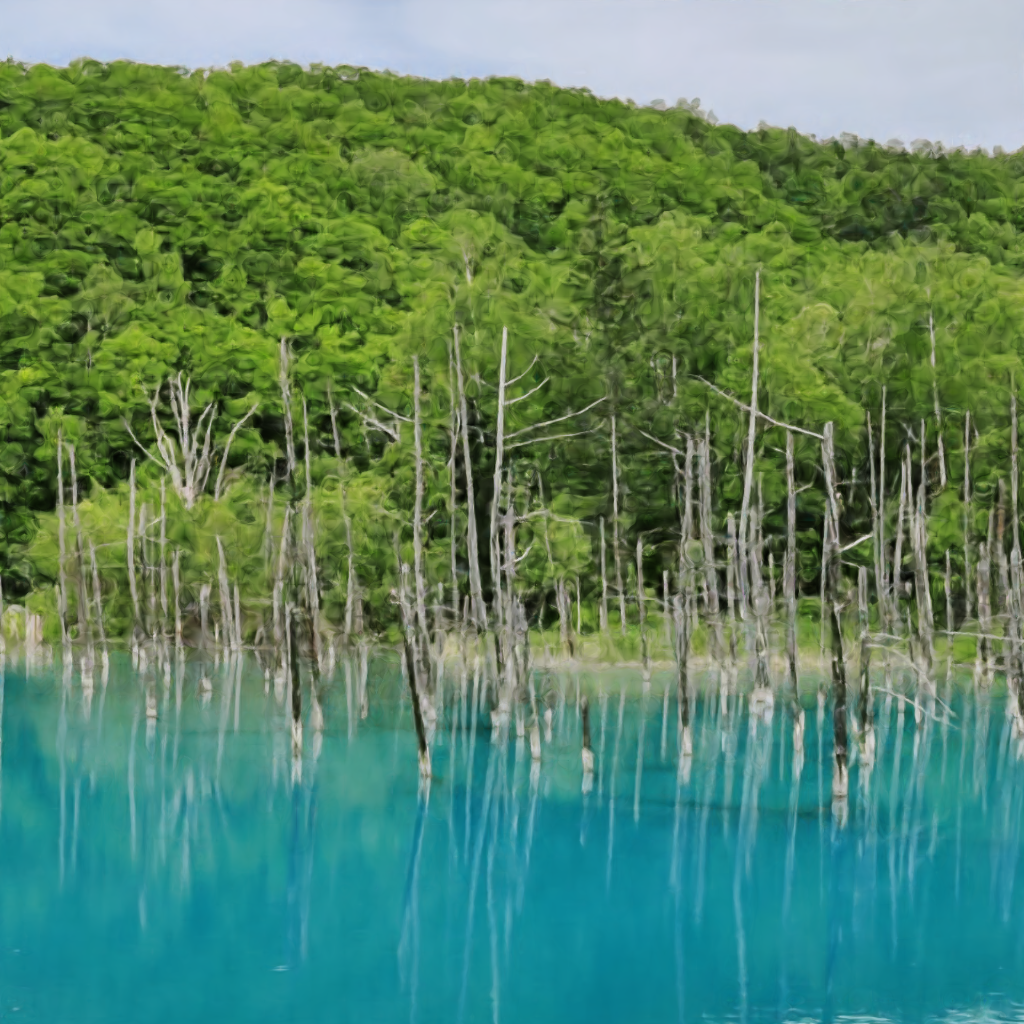} &
        \includegraphics[width=0.24\linewidth]{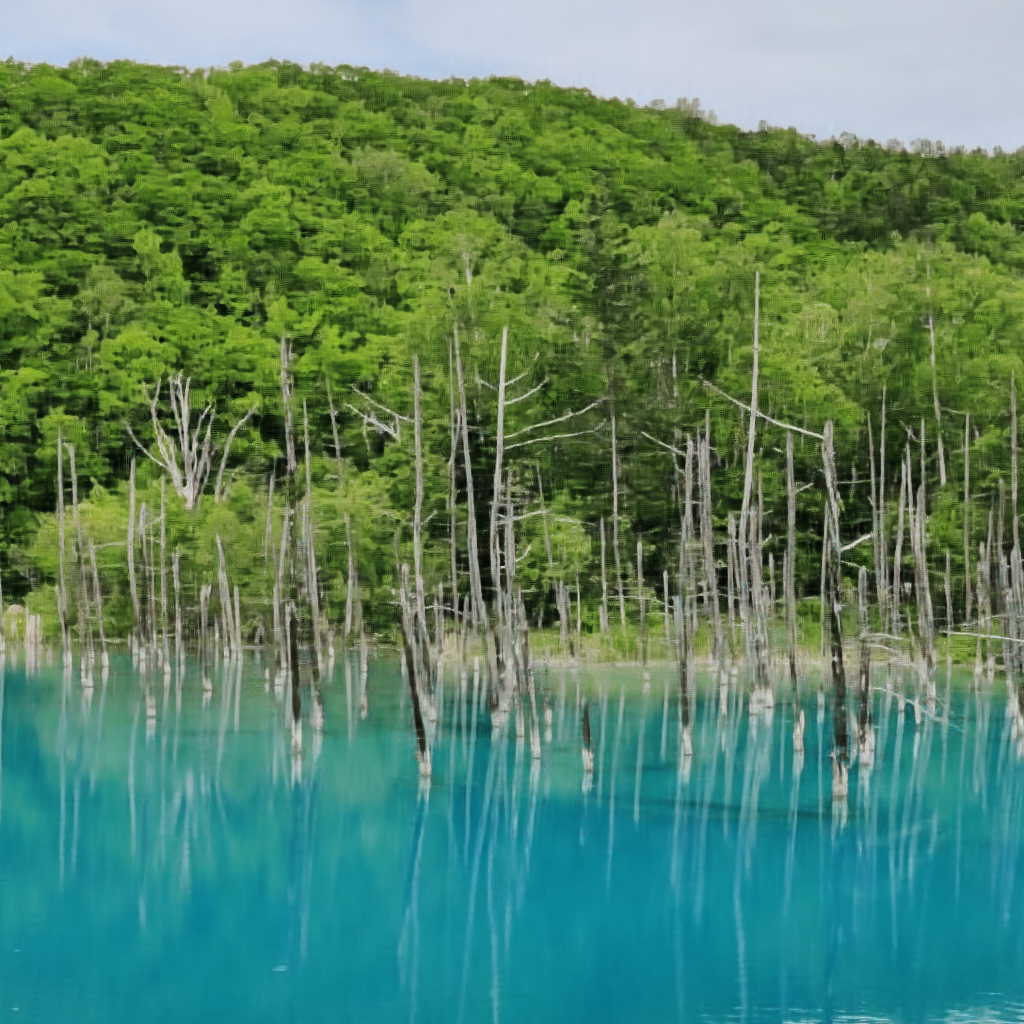} &
        \includegraphics[width=0.24\linewidth]{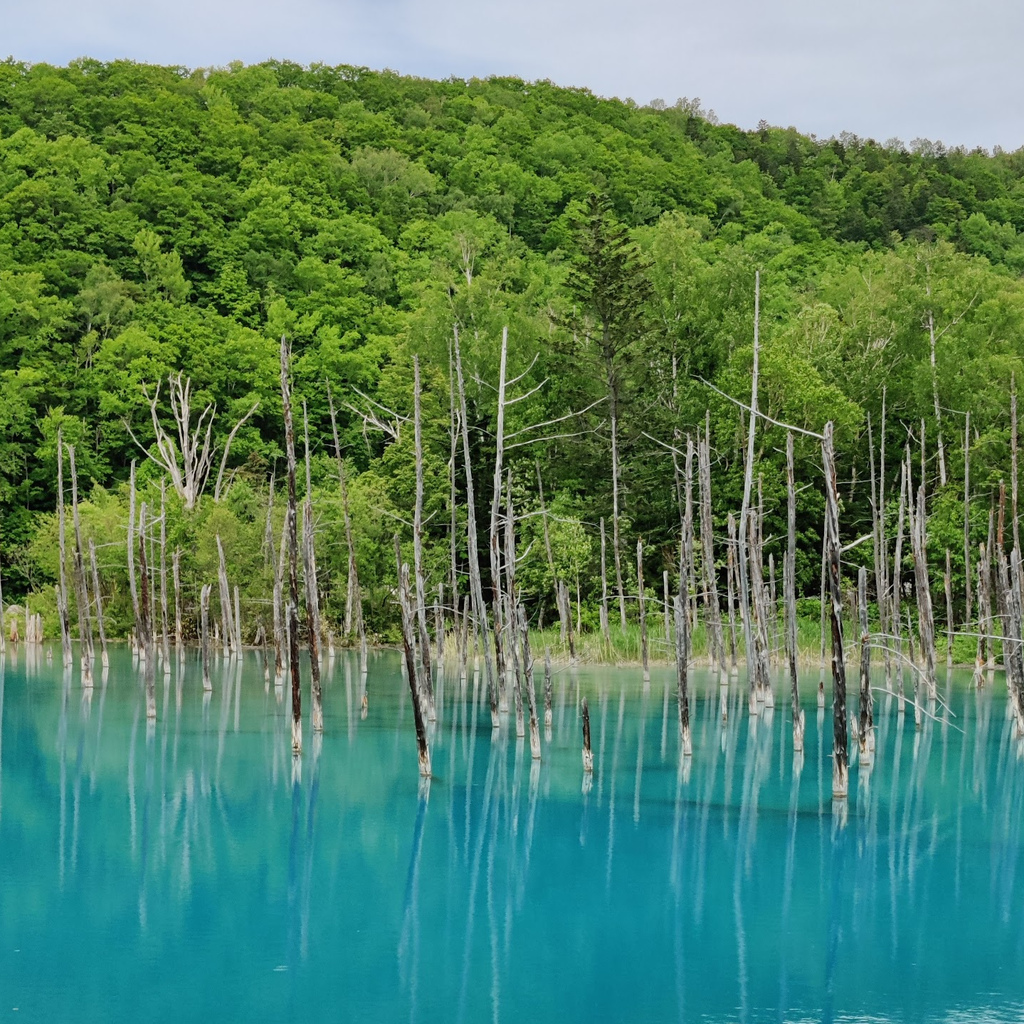} \\
        
        & \small{(a) Positional Encoding} & \small{(b) Grid Encoding} & \small{(c) Local Positional Encoding} & \small{(d) Reference}
    \end{tabular}
    \caption{Qualitative comparison among different encodings for an MLP. They all use a $64\times64$ grid with $16$-dimensional latent vectors and $4$ frequencies in positional encoding. The MLP has three hidden layers with $64$ neurons each. The resolutions of all images are $1,024 \times 1,024$.}
    \label{fig:table}
\end{figure*}

In this task, we train the network to map a two-dimensional input coordinate to the RGB color of an image at that location.
Fig.~\ref{fig:teaser} and Fig.~\ref{fig:table} compare the images reconstructed using the MLP with different input encodings, which are trained for $1$k iterations.
To make the comparison fair, we used the same condition for all the encodings.
Local positional encoding uses a grid with $N = 64$ and positional encoding with $n = 4$.
We choose these parameters experimentally. A parameter study can be found in the supplemental document.
For the equivalent memory consumption, grid encoding also uses a grid with $N = 64$ which stores a $16$-dimensional latent vector in each cell, and positional encoding uses $4$ frequencies.
All these encodings result in a $16$-dimensional input vector to the MLP. 
All the images we used for training are $1,024\times1,024$ resolution and the network reconstructs images with the same resolution.
As illustrated in Fig.~\ref{fig:teaser} and Fig.~\ref{fig:table}, we can see that local positional encoding produces visually finer results for all images compared to other encodings while we can clearly observe that the results from grid encoding and positional encoding cannot capture high-frequency details.
More detailed comparisons with close-up images can be found in Fig.~\ref{fig:teaser} and Fig.~\ref{fig:zoomup}.
Additionally, Table~\ref{tab:image_quantitative} shows the quantitative comparisons with PSNR and SSIM~\cite{1284395} where a higher value indicates better quality.
Local positional encoding achieves the highest values in both PSNR and SSIM for all images.

\begin{figure}
    \centering
    \setlength{\tabcolsep}{0\linewidth}
    \renewcommand{\arraystretch}{0}
    \begin{tabular}{cccc}
        \multirow[c]{3}{*}[1.03cm]{ \begin{overpic}[width=0.375\linewidth]{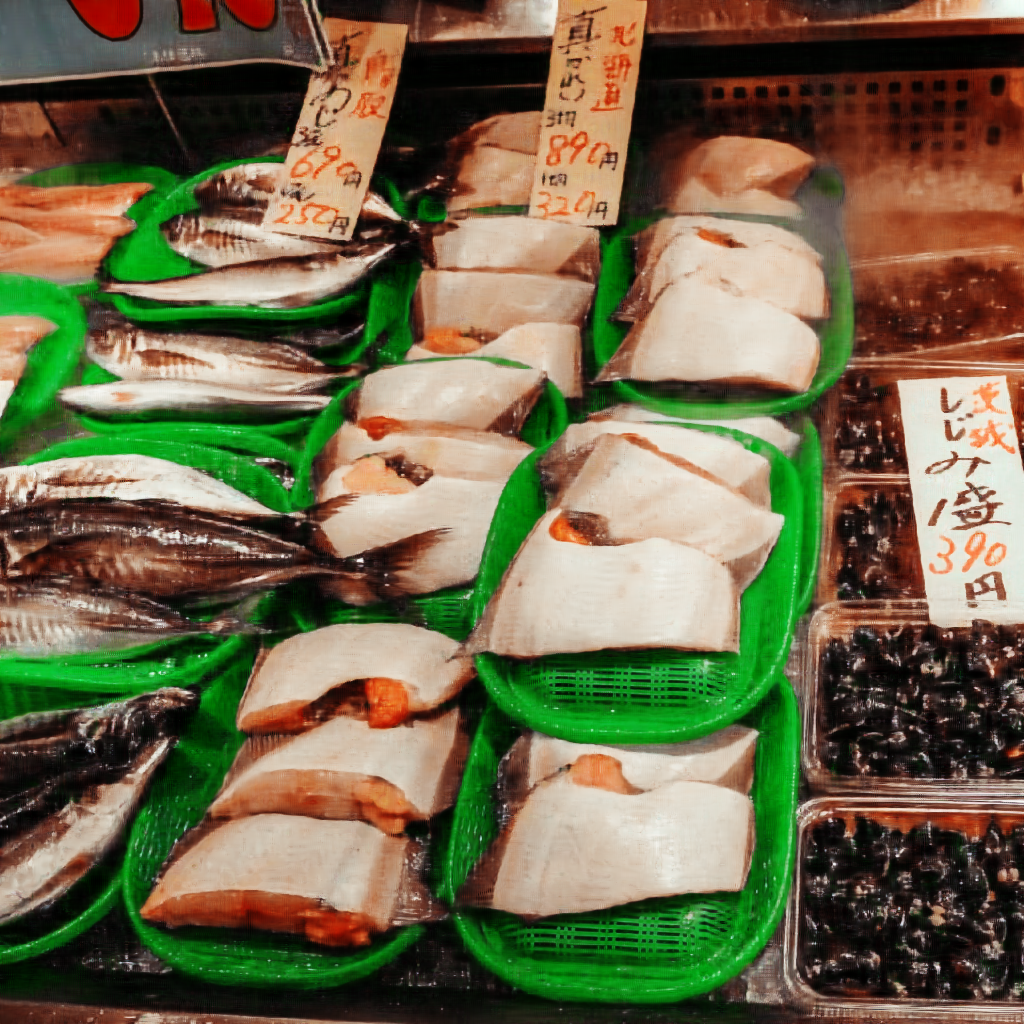} 
            \put(520,872){\linethickness{0.3mm}\color{red}\polygon(0,0)(118,0)(118,118)(0,118)}
            \put(252,762){\linethickness{0.3mm}\color{blue}\polygon(0,0)(118,0)(118,118)(0,118)}
            \put(510,264){\linethickness{0.3mm}\color{green}\polygon(0,0)(118,0)(118,118)(0,118)}
        \end{overpic} } &
        \includegraphics[width=0.116\linewidth, viewport=533 893 651 1011, clip, cfbox=red 1pt 0pt]{figs/sameCoeffDiff/auto_fish_lpe1_noIdx.png} &
        \includegraphics[width=0.116\linewidth, viewport=533 893 651 1011, clip, cfbox=red 1pt 0pt]{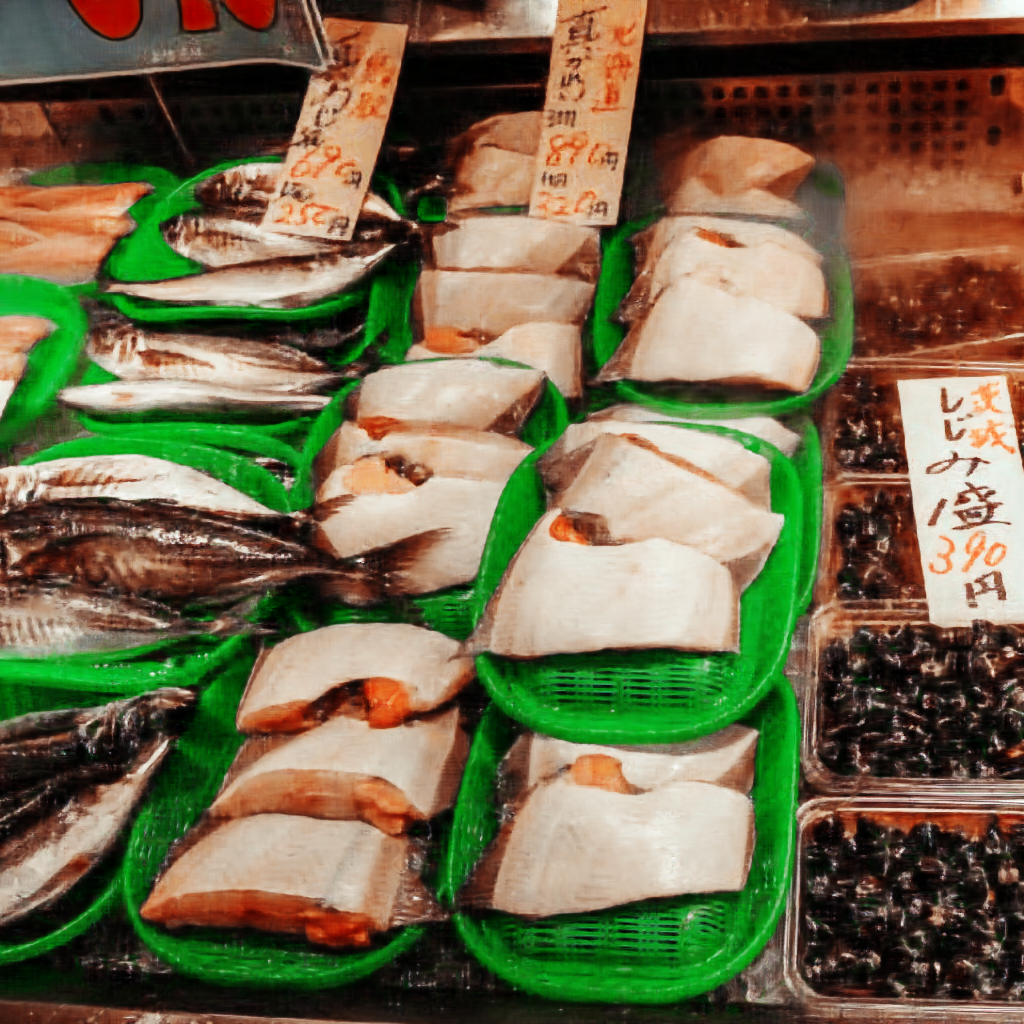} &
        \multirow[c]{3}{*}[1.03cm]{ \begin{overpic}[width=0.375\linewidth]{figs/sameCoeffDiff/auto_fish_lpe1_sameCoef_noIdx.png} 
            \put(520,872){\linethickness{0.3mm}\color{red}\polygon(0,0)(118,0)(118,118)(0,118)}
            \put(252,762){\linethickness{0.3mm}\color{blue}\polygon(0,0)(118,0)(118,118)(0,118)}
            \put(510,264){\linethickness{0.3mm}\color{green}\polygon(0,0)(118,0)(118,118)(0,118)}
        \end{overpic} } \\

        & \includegraphics[width=0.116\linewidth, viewport=258 780 376 898, clip, cfbox=blue 1pt 0pt]{figs/sameCoeffDiff/auto_fish_lpe1_noIdx.png} &
        \includegraphics[width=0.116\linewidth, viewport=258 780 376 898, clip, cfbox=blue 1pt 0pt]{figs/sameCoeffDiff/auto_fish_lpe1_sameCoef_noIdx.png} & \\

        & \includegraphics[width=0.116\linewidth, viewport=513 270 631 388, clip, cfbox=green 1pt 0pt]{figs/sameCoeffDiff/auto_fish_lpe1_noIdx.png} &
        \includegraphics[width=0.116\linewidth, viewport=513 270 631 388, clip, cfbox=green 1pt 0pt]{figs/sameCoeffDiff/auto_fish_lpe1_sameCoef_noIdx.png} & \\

        \addlinespace
        \small{PSNR: 24.03, SSIM: 0.7071}& & & \small{PSNR: 23.28, SSIM: 0.6776} \\

        \addlinespace
        \small{Different coefficients }&  &  & \small{Same coefficients }
    \end{tabular}
    \caption{ Comparison of \textsc{Fish Market} images reconstructed using the different coefficients (Left), and the same coefficients which halve the number of parameters (Right), for sinusoidal functions. }
    \label{fig:sameCoeff}
\end{figure}

Note that we also tried to use the same coefficient for sinusoidal functions with the same frequency (i.e. $a^c_i = a^s_i = a_i$) in order to decrease memory consumption for the grid of local positional encoding.
However, this trades off the image quality for memory consumption.
As shown in Fig.~\ref{fig:sameCoeff}, using the same coefficients decreases memory usage by roughly half, but it leads to worse results.
In this paper, we recommend that the different coefficients are used for each sinusoidal function as discussed in Sec.~\ref{Sec:lpe}, but the optimal selection of approaches depends on the specific use case.

\subsection{Signed Distance Functions} \label{sec:sdf}
\begin{figure*}
    \centering
    \setlength{\tabcolsep}{0.002\linewidth}
    \begin{tabular}{ccccc}
        \raisebox{0.11\linewidth}{\rotatebox[origin = c]{90}{\textsc{Thai Statue}}} &
        \begin{overpic}[width=0.24\linewidth]{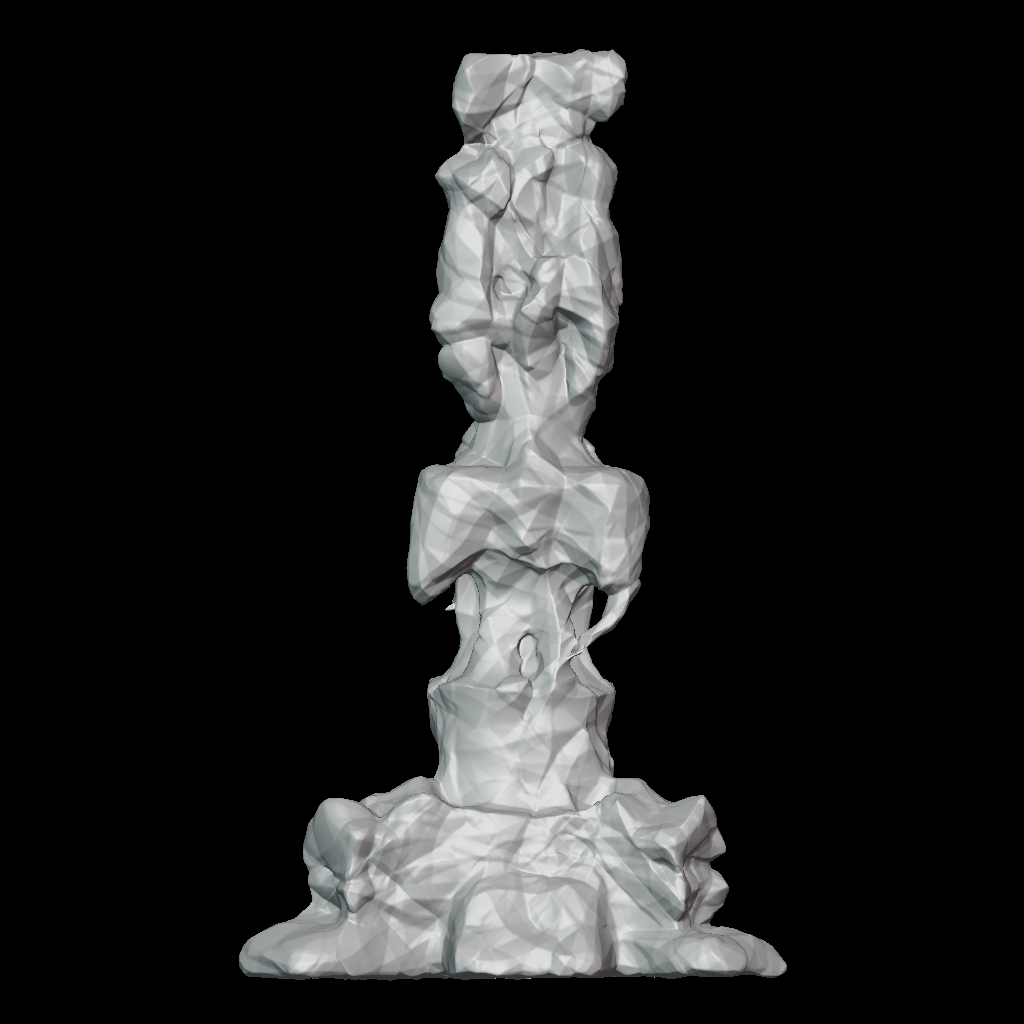}
            \put(443, 659){ \linethickness{0.1mm}\color{red}\polygon(0,0)(99,0)(99,99)(0,99)}
            \put(730, 255){ \includegraphics[width=0.058\linewidth,viewport=474 676 574 776, clip, cfbox=red 0.5pt 0pt]{figs/sdf/cmp3encs_2x/auto_sdf_POSITIONAL_thaiStatue_512_16384.png} }
            \put(441, 224){ \linethickness{0.1mm}\color{blue}\polygon(0,0)(99,0)(99,99)(0,99)}
            \put(730, 5){ \includegraphics[width=0.058\linewidth,viewport=469 231 569 331, clip, cfbox=blue 0.5pt 0pt]{figs/sdf/cmp3encs_2x/auto_sdf_POSITIONAL_thaiStatue_512_16384.png} }
        \end{overpic} &
        \begin{overpic}[width=0.24\linewidth]{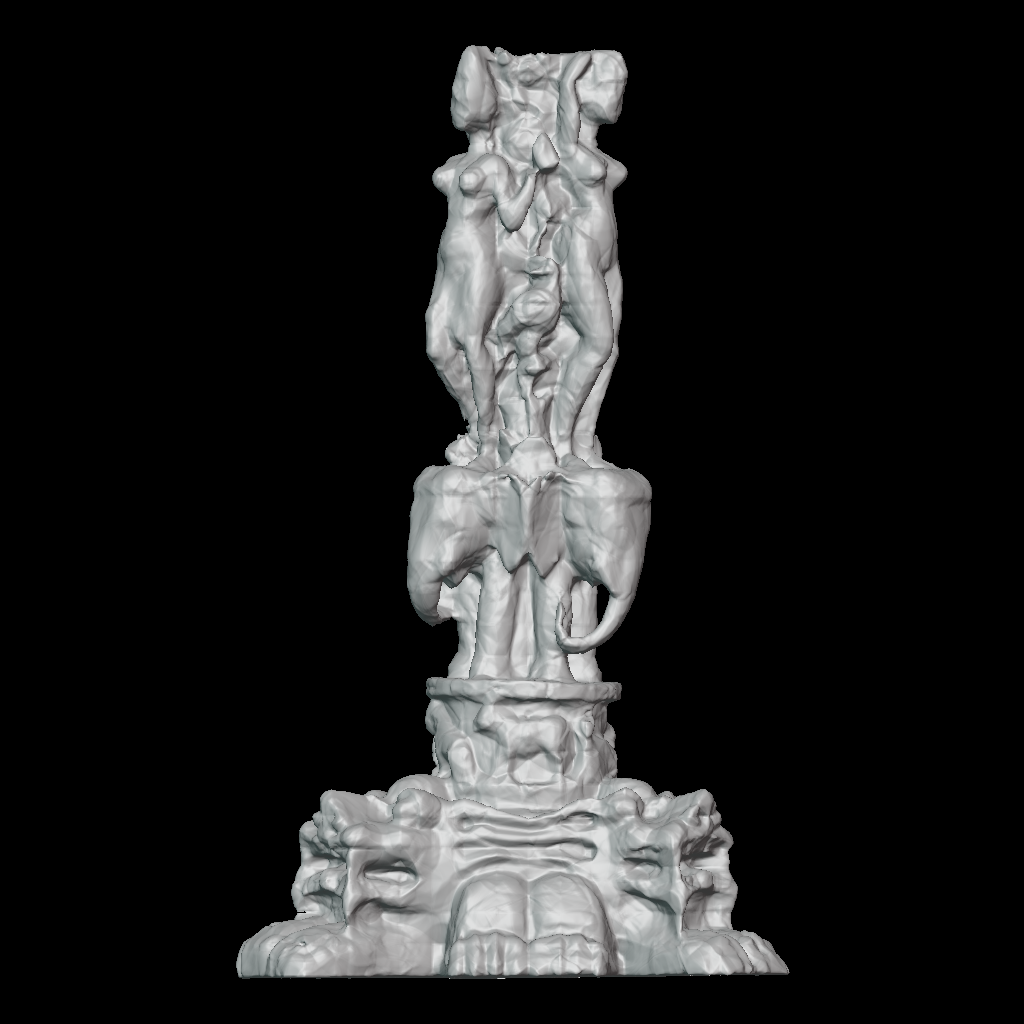} 
            \put(443, 659){ \linethickness{0.1mm}\color{red}\polygon(0,0)(99,0)(99,99)(0,99)}
            \put(730, 255){ \includegraphics[width=0.058\linewidth,viewport=474 676 574 776, clip, cfbox=red 0.5pt 0pt]{figs/sdf/cmp3encs_2x/auto_sdf_GRID_thaiStatue_512_16384.png} }
            \put(441, 224){ \linethickness{0.1mm}\color{blue}\polygon(0,0)(99,0)(99,99)(0,99)}
            \put(730, 5){ \includegraphics[width=0.058\linewidth,viewport=469 231 569 331, clip, cfbox=blue 0.5pt 0pt]{figs/sdf/cmp3encs_2x/auto_sdf_GRID_thaiStatue_512_16384.png} }
        \end{overpic}&
        \begin{overpic}[width=0.24\linewidth]{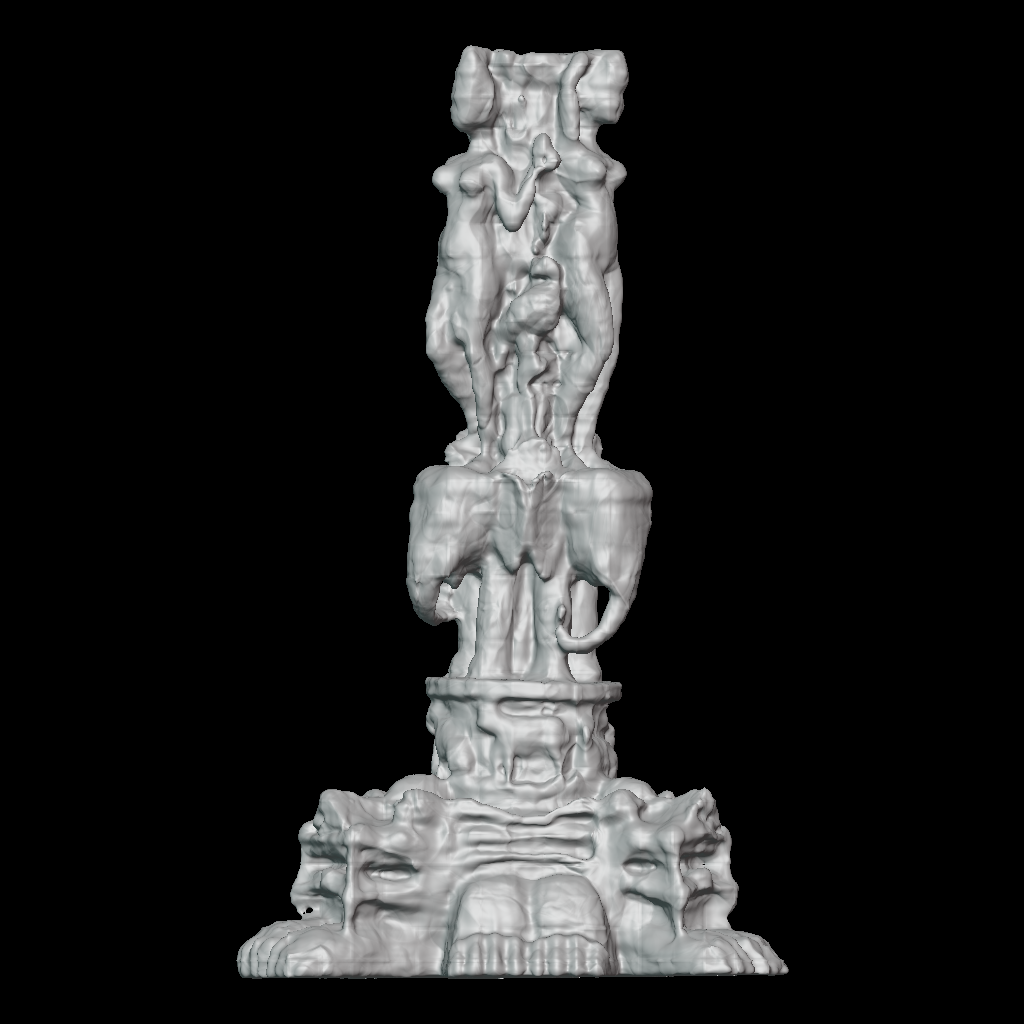} 
            \put(443, 659){ \linethickness{0.1mm}\color{red}\polygon(0,0)(99,0)(99,99)(0,99)}
            \put(730, 255){ \includegraphics[width=0.058\linewidth,viewport=474 676 574 776, clip, cfbox=red 0.5pt 0pt]{figs/sdf/cmp3encs_2x/auto_sdf_LPE_thaiStatue_512_itr16384_A.png} }
            \put(441, 224){ \linethickness{0.1mm}\color{blue}\polygon(0,0)(99,0)(99,99)(0,99)}
            \put(730, 5){ \includegraphics[width=0.058\linewidth,viewport=469 231 569 331, clip, cfbox=blue 0.5pt 0pt]{figs/sdf/cmp3encs_2x/auto_sdf_LPE_thaiStatue_512_itr16384_A.png} }
        \end{overpic} &
        \begin{overpic}[width=0.24\linewidth]{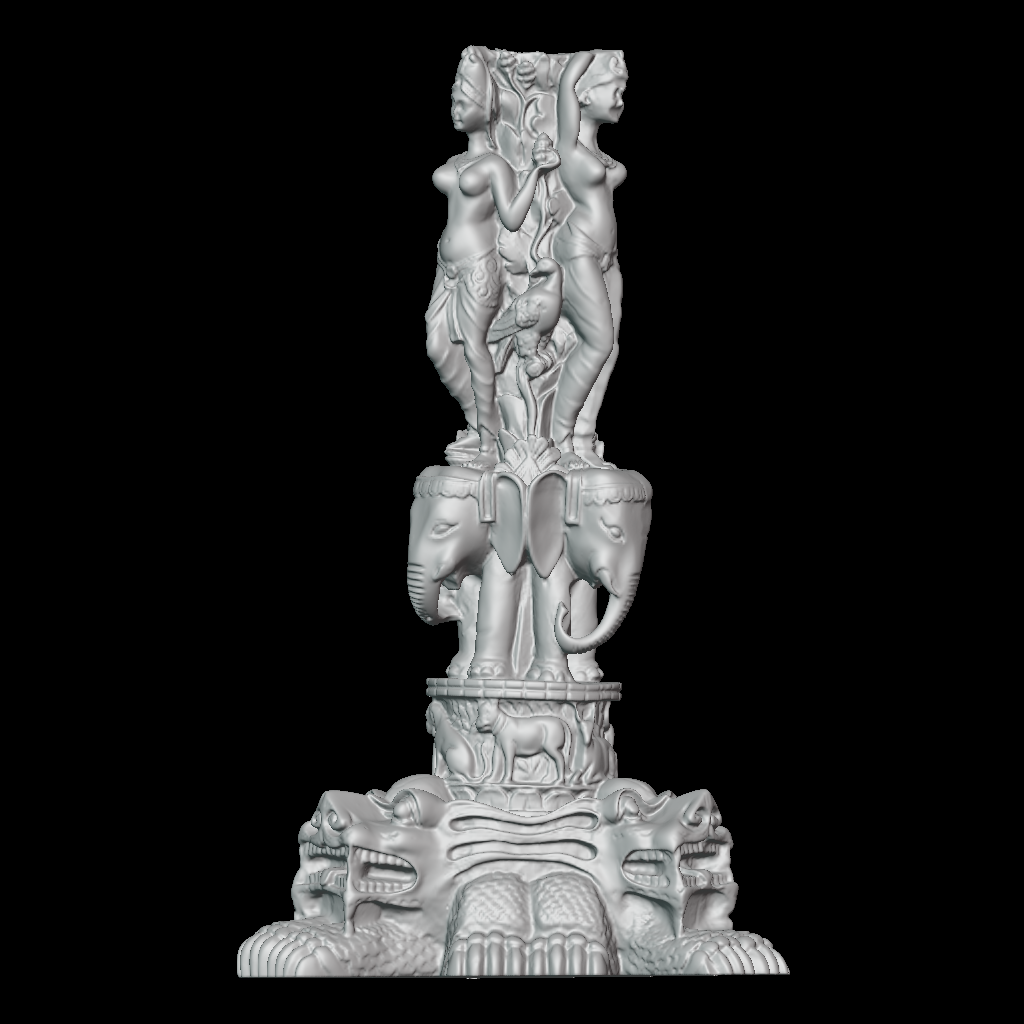} 
            \put(443, 659){ \linethickness{0.1mm}\color{red}\polygon(0,0)(99,0)(99,99)(0,99)}
            \put(730, 255){ \includegraphics[width=0.058\linewidth,viewport=474 676 574 776, clip, cfbox=red 0.5pt 0pt]{figs/sdf/sdf_thai_512_ref.png} }
            \put(441, 224){ \linethickness{0.1mm}\color{blue}\polygon(0,0)(99,0)(99,99)(0,99)}
            \put(730, 5){ \includegraphics[width=0.058\linewidth,viewport=469 231 569 331, clip, cfbox=blue 0.5pt 0pt]{figs/sdf/sdf_thai_512_ref.png} }
        \end{overpic} \\

        & IoU = 0.9065 & \textbf{ IoU = 0.9755 } & IoU = 0.9736 \\
        \addlinespace

        & \small{(a) Positional Encoding} & \small{(b) Grid Encoding} & \small{(c) Local Positional Encoding } & \small{(d) Reference}
    \end{tabular}
    \caption{ Qualitative and quantitative comparison with positional encoding, grid encoding, and local positional encoding with SDFs geometry rendering. The shading is applied to show surface details by the Lit Sphere~\cite{LitSphere} based on its geometric normal. They all use a $32\times32\times32$ grid with $18$-dimensional latent vectors and $3$ frequencies in positional encoding. The MLP has three hidden layers with $64$ neurons each. Each close-up view enclosed by red and blue squares represents how the fine details are captured by the encoding. IoU metrics are shown for each encoding. The bold number represents the best in the encodings for the geometry.} 
    \label{fig:imagesdf}
\end{figure*}

\begin{figure*}
    \centering
    \setlength{\tabcolsep}{0\linewidth}
    \def\arraystretch{0.0}
    \begin{tabular}{c@{\hspace{0.002\linewidth}} c c@{\hspace{0.002\linewidth}} c c@{\hspace{0.002\linewidth}} c c}
        \multirow[c]{3}{*}[0cm]{ \raisebox{0.14\linewidth}{\rotatebox[origin = c]{90}{Grid Encoding}} } &
        \multirow[c]{3}{*}[1.71cm]{ \begin{overpic}[width=0.18\linewidth, trim=200 0 200 0, clip]{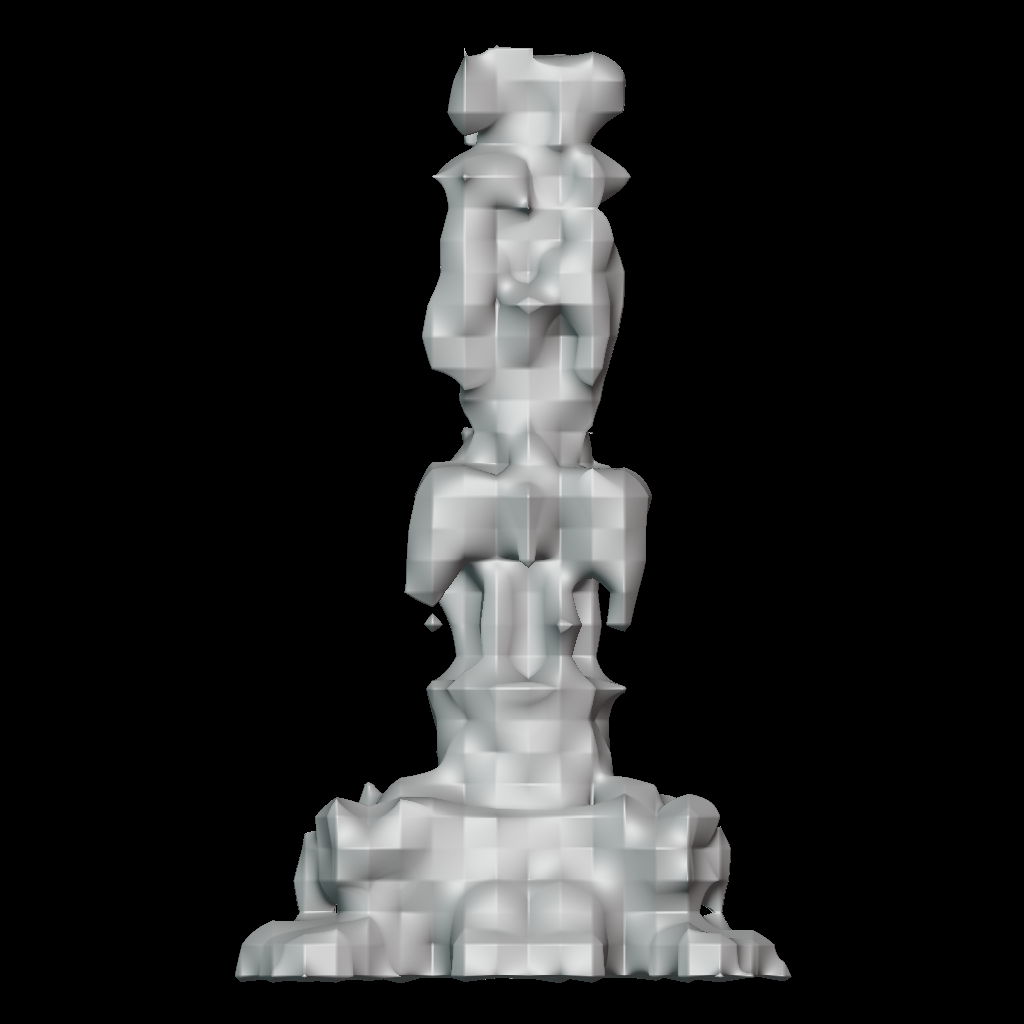} 
            \put(224,36){\linethickness{0.2mm}\color{green}\polygon(0,0)(100,0)(100,100)(0,100)}
            \put(338,357){\linethickness{0.2mm}\color{blue}\polygon(0,0)(100,0)(100,100)(0,100)}
            \put(323,859){\linethickness{0.2mm}\color{red}\polygon(0,0)(100,0)(100,100)(0,100)}
        \end{overpic} } &
        \includegraphics[width=0.095\linewidth,viewport=536 879 636 979, clip, cfbox=red 1pt 0pt]{figs/sdf/cmpConv_2x/auto_sdf_GRID_thaiStatue_512_64.png} &
        \multirow[c]{3}{*}[1.71cm]{ \begin{overpic}[width=0.18\linewidth, trim=200 0 200 0, clip]{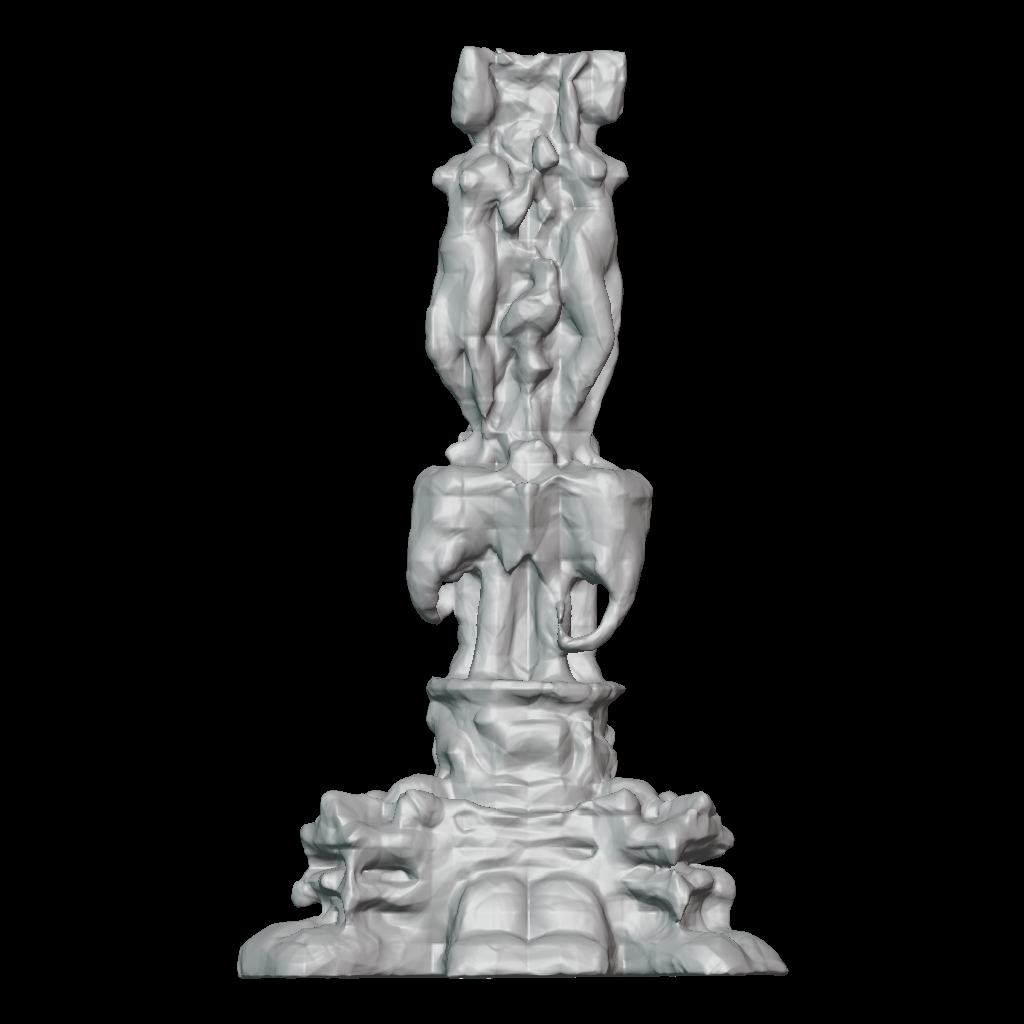}
            \put(224,36){\linethickness{0.2mm}\color{green}\polygon(0,0)(100,0)(100,100)(0,100)}
            \put(338,357){\linethickness{0.2mm}\color{blue}\polygon(0,0)(100,0)(100,100)(0,100)}
            \put(323,859){\linethickness{0.2mm}\color{red}\polygon(0,0)(100,0)(100,100)(0,100)}
        \end{overpic} } &
        \includegraphics[width=0.095\linewidth,viewport=536 879 636 979, clip, cfbox=red 1pt 0pt]{figs/sdf/cmpConv_2x/auto_sdf_GRID_thaiStatue_512_1024.png} &
        \multirow[c]{3}{*}[1.71cm]{ \begin{overpic}[width=0.18\linewidth, trim=200 0 200 0, clip]{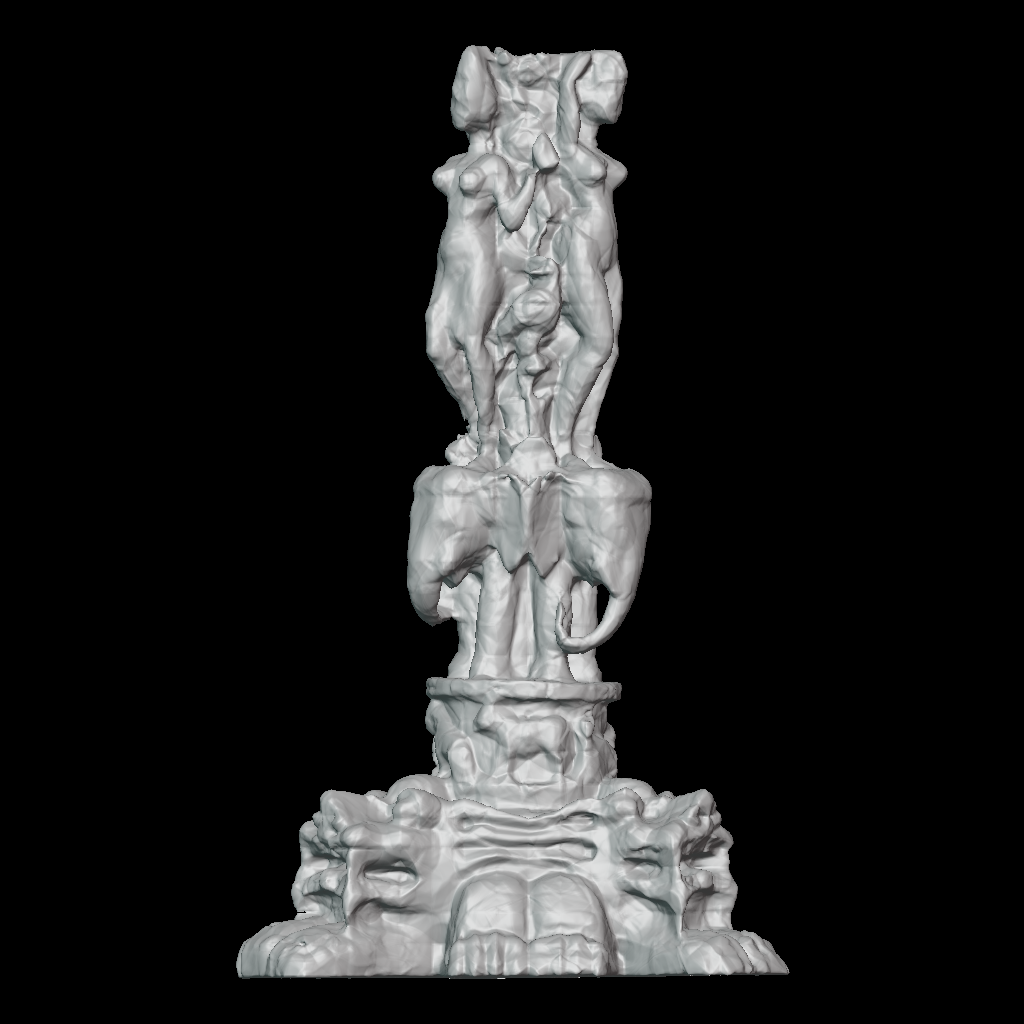}
            \put(224,36){\linethickness{0.2mm}\color{green}\polygon(0,0)(100,0)(100,100)(0,100)}
            \put(338,357){\linethickness{0.2mm}\color{blue}\polygon(0,0)(100,0)(100,100)(0,100)}
            \put(323,859){\linethickness{0.2mm}\color{red}\polygon(0,0)(100,0)(100,100)(0,100)}
        \end{overpic} } &
        \includegraphics[width=0.095\linewidth,viewport=536 879 636 979, clip, cfbox=red 1pt 0pt]{figs/sdf/cmpConv_2x/auto_sdf_GRID_thaiStatue_512_16384.png} \\

        & & \includegraphics[width=0.095\linewidth,viewport=552 367 652 467, clip, cfbox=blue 1pt 0pt]{figs/sdf/cmpConv_2x/auto_sdf_GRID_thaiStatue_512_64.png} &
        & \includegraphics[width=0.095\linewidth,viewport=552 367 652 467, clip, cfbox=blue 1pt 0pt]{figs/sdf/cmpConv_2x/auto_sdf_GRID_thaiStatue_512_1024.png} &
        & \includegraphics[width=0.095\linewidth,viewport=552 367 652 467, clip, cfbox=blue 1pt 0pt]{figs/sdf/cmpConv_2x/auto_sdf_GRID_thaiStatue_512_16384.png} \\

        & & \includegraphics[width=0.095\linewidth,viewport=435 38 535 138, clip, cfbox=green 1pt 0pt]{figs/sdf/cmpConv_2x/auto_sdf_GRID_thaiStatue_512_64.png} &
        & \includegraphics[width=0.095\linewidth,viewport=435 38 535 138, clip, cfbox=green 1pt 0pt]{figs/sdf/cmpConv_2x/auto_sdf_GRID_thaiStatue_512_1024.png} &
        & \includegraphics[width=0.095\linewidth,viewport=435 38 535 138, clip, cfbox=green 1pt 0pt]{figs/sdf/cmpConv_2x/auto_sdf_GRID_thaiStatue_512_16384.png} \\

        \addlinespace

        \multirow[c]{3}{*}[0.8cm]{ \raisebox{0.14\linewidth}{\rotatebox[origin = c]{90}{Local Positional Encoding}} } &
        \multirow[c]{3}{*}[1.71cm]{ \begin{overpic}[width=0.18\linewidth, trim=200 0 200 0, clip]{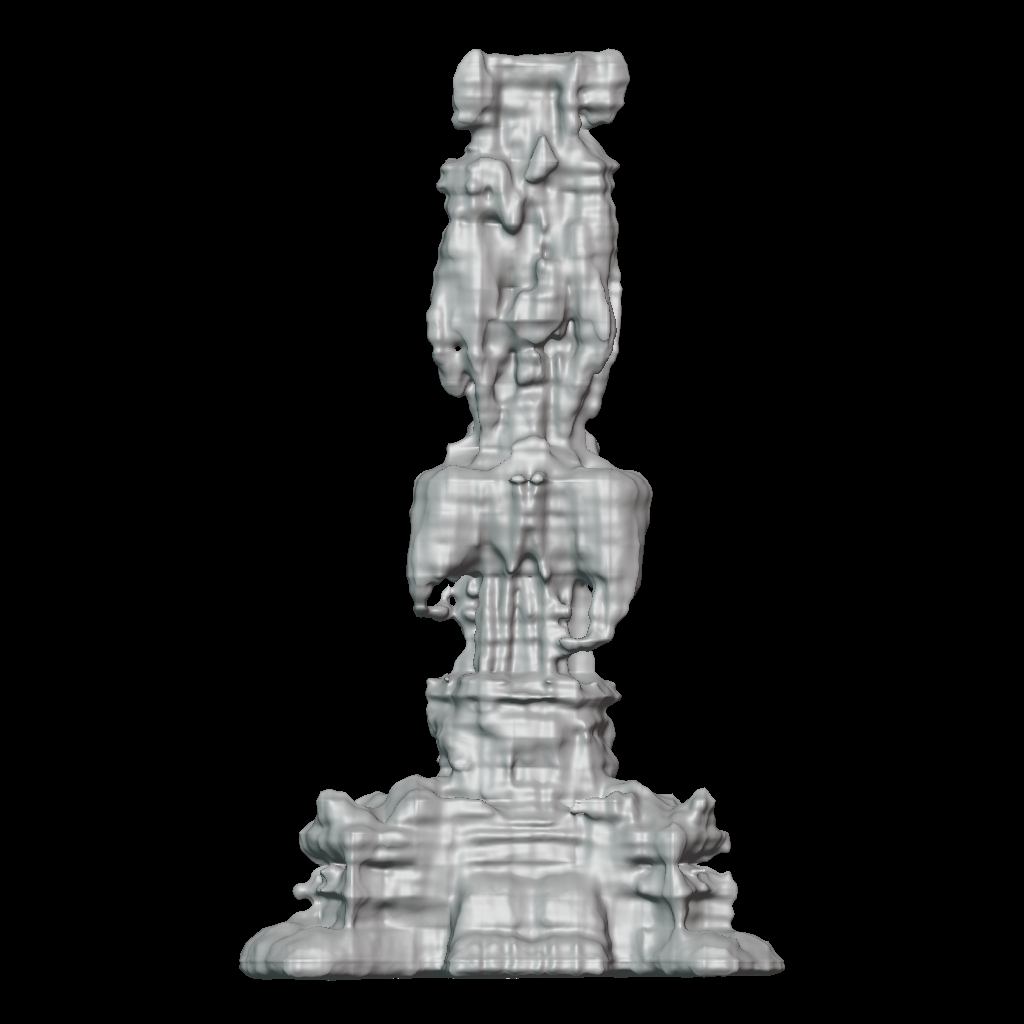} 
            \put(224,36){\linethickness{0.2mm}\color{green}\polygon(0,0)(100,0)(100,100)(0,100)}
            \put(338,357){\linethickness{0.2mm}\color{blue}\polygon(0,0)(100,0)(100,100)(0,100)}
            \put(323,859){\linethickness{0.2mm}\color{red}\polygon(0,0)(100,0)(100,100)(0,100)}
        \end{overpic} } &
        \includegraphics[width=0.095\linewidth,viewport=536 879 636 979, clip, cfbox=red 1pt 0pt]{figs/sdf/cmpConv_2x/auto_sdf_LPE_thaiStatue_512_itr64_A.png} &
        \multirow[c]{3}{*}[1.71cm]{ \begin{overpic}[width=0.18\linewidth, trim=200 0 200 0, clip]{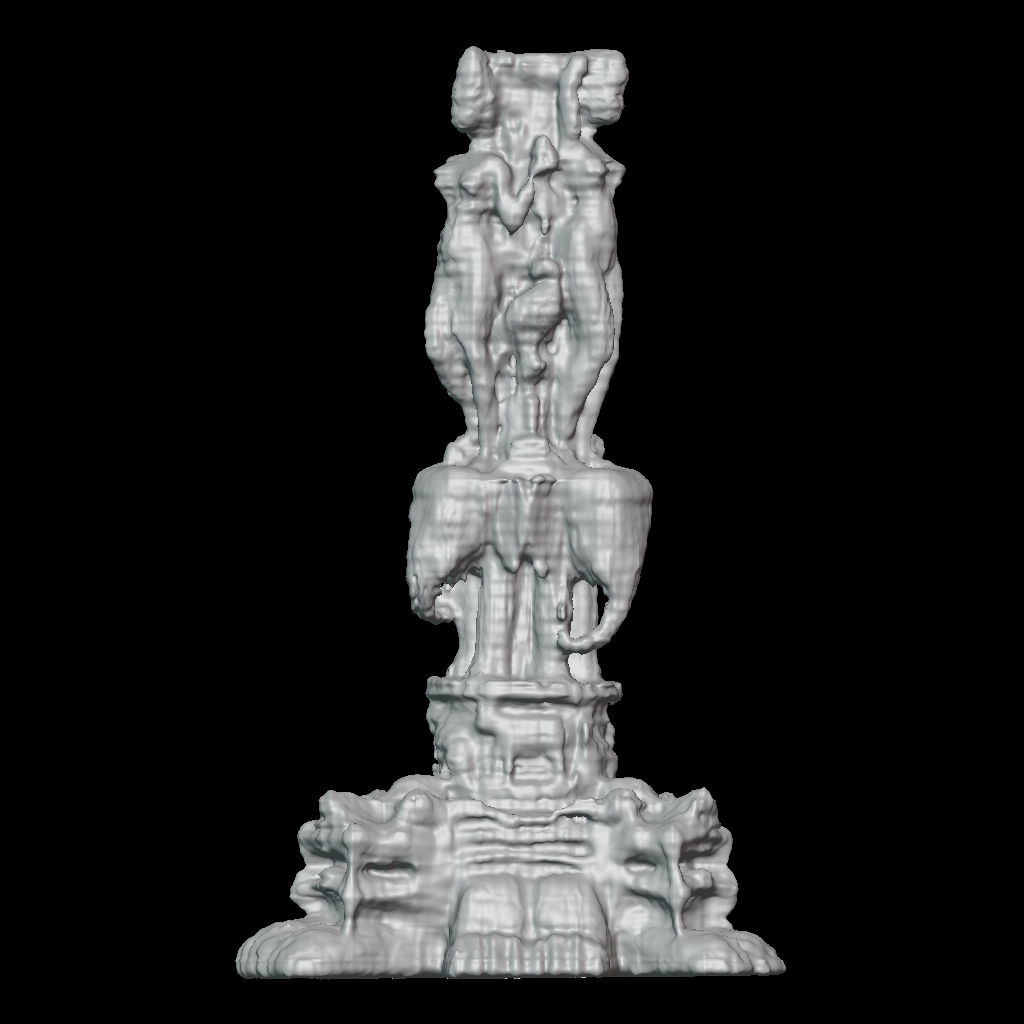}
            \put(224,36){\linethickness{0.2mm}\color{green}\polygon(0,0)(100,0)(100,100)(0,100)}
            \put(338,357){\linethickness{0.2mm}\color{blue}\polygon(0,0)(100,0)(100,100)(0,100)}
            \put(323,859){\linethickness{0.2mm}\color{red}\polygon(0,0)(100,0)(100,100)(0,100)}
        \end{overpic} } &
        \includegraphics[width=0.095\linewidth,viewport=536 879 636 979, clip, cfbox=red 1pt 0pt]{figs/sdf/cmpConv_2x/auto_sdf_LPE_thaiStatue_512_itr1024_A.png} &
        \multirow[c]{3}{*}[1.71cm]{ \begin{overpic}[width=0.18\linewidth, trim=200 0 200 0, clip]{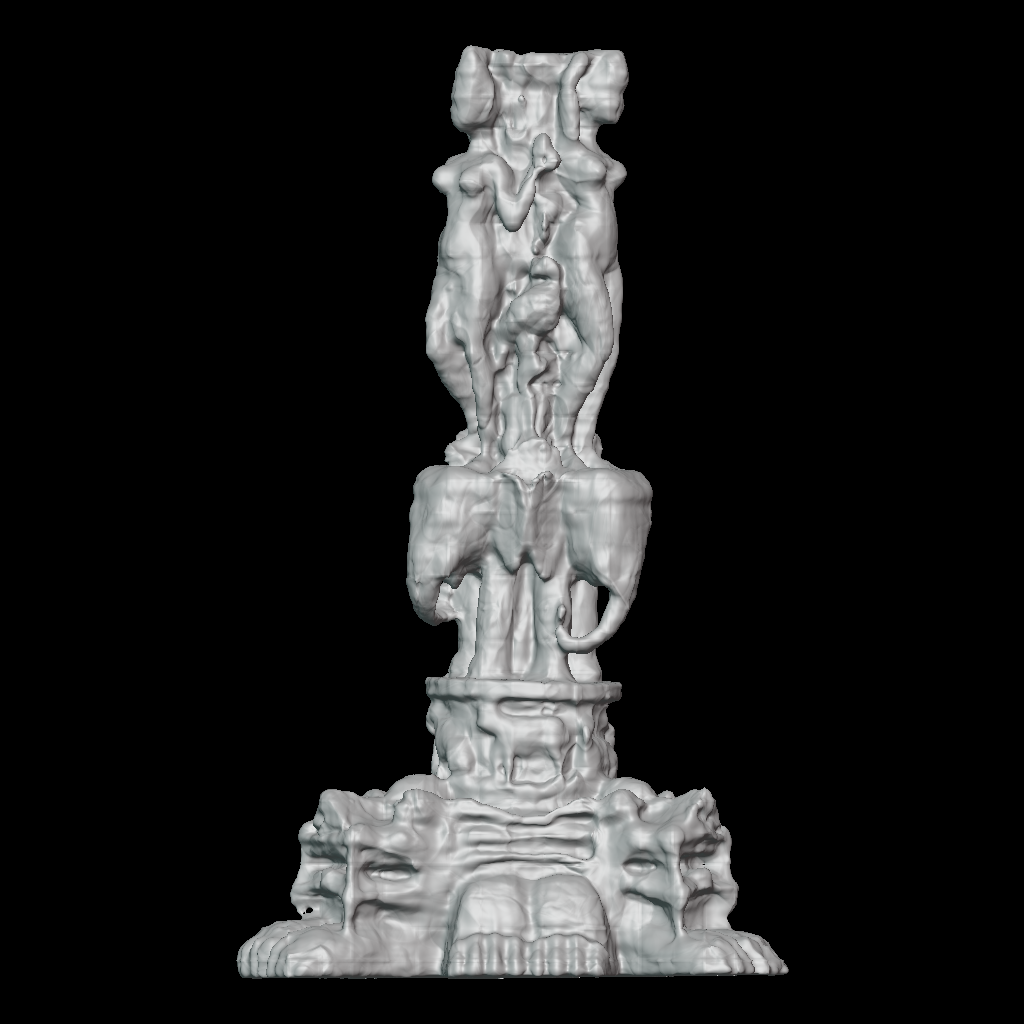}
            \put(224,36){\linethickness{0.2mm}\color{green}\polygon(0,0)(100,0)(100,100)(0,100)}
            \put(338,357){\linethickness{0.2mm}\color{blue}\polygon(0,0)(100,0)(100,100)(0,100)}
            \put(323,859){\linethickness{0.2mm}\color{red}\polygon(0,0)(100,0)(100,100)(0,100)}
        \end{overpic} } &
        \includegraphics[width=0.095\linewidth,viewport=536 879 636 979, clip, cfbox=red 1pt 0pt]{figs/sdf/cmpConv_2x/auto_sdf_LPE_thaiStatue_512_itr16384_A.png} \\

        & & \includegraphics[width=0.095\linewidth,viewport=552 367 652 467, clip, cfbox=blue 1pt 0pt]{figs/sdf/cmpConv_2x/auto_sdf_LPE_thaiStatue_512_itr64_A.png} &
        & \includegraphics[width=0.095\linewidth,viewport=552 367 652 467, clip, cfbox=blue 1pt 0pt]{figs/sdf/cmpConv_2x/auto_sdf_LPE_thaiStatue_512_itr1024_A.png} &
        & \includegraphics[width=0.095\linewidth,viewport=552 367 652 467, clip, cfbox=blue 1pt 0pt]{figs/sdf/cmpConv_2x/auto_sdf_LPE_thaiStatue_512_itr16384_A.png} \\

        & & \includegraphics[width=0.095\linewidth,viewport=435 38 535 138, clip, cfbox=green 1pt 0pt]{figs/sdf/cmpConv_2x/auto_sdf_LPE_thaiStatue_512_itr64_A.png} &
        & \includegraphics[width=0.095\linewidth,viewport=435 38 535 138, clip, cfbox=green 1pt 0pt]{figs/sdf/cmpConv_2x/auto_sdf_LPE_thaiStatue_512_itr1024_A.png} &
        & \includegraphics[width=0.095\linewidth,viewport=435 38 535 138, clip, cfbox=green 1pt 0pt]{figs/sdf/cmpConv_2x/auto_sdf_LPE_thaiStatue_512_itr16384_A.png} \\
        
        \addlinespace

        & \multicolumn{2}{c}{ \small{(a) 64 training iterations } } & 
        \multicolumn{2}{c}{ \small{(b) 1,024 training iterations } } & 
        \multicolumn{2}{c}{ \small{(c) 16,384 training iterations } } \\
    \end{tabular}
    \caption{ Convergence comparisons with different training iterations between grid encoding and local positional encoding with the \textsc{Thai Statue} model. Our method shows finer geometric details in the earlier training stage. }
    \label{fig:imagesdfconverge}
\end{figure*}

Signed Distance Functions (SDFs) define the closest distance to surfaces for any spatial location to represent a shape.
SDFs can be defined by discretized grid representation; however, they require a lot of memory for dense grids. 
We evaluate our method by letting the network learn three-dimensional SDFs and comparing the accuracy of the reconstructed SDFs with a small grid resolution. 
Local positional encoding uses a grid with $N = 32$ and positional encoding with $n = 3$ which results in an $18$-dimensional input vector to the MLP.
As with the image reconstruction task, a parameter study for SDFs also can be found in the supplemental document.
For a fair comparison, grid encoding also uses a grid with $N = 32$ which stores an $18$-dimensional latent vector in each cell to make the memory consumption equivalent, and positional encoding uses $3$ frequencies to align the input vector dimension.
As a qualitative evaluation, we render a shape by SDFs with shading to emphasize high-frequency differences of the surface among the encodings. We use the Lit Sphere~\cite{LitSphere} for shading for consistent and reproducible results.
Fig.~\ref{fig:imagesdf} shows rendered images with the \textsc{Thai Statue} model with different encodings along with the intersection-over-union (IoU) metric for a quantitative comparison, which is a ratio between the intersection and union of two volumes. 
We calculate IoUs by taking a sign of $128$ million sampling points around the bounding box of the shape on the reference SDFs.
Comparisons with more geometries can be found in the supplemental document.
The close-up views of our method (Fig.~\ref{fig:imagesdf}c) show high-frequency details well while grid encoding (Fig.~\ref{fig:imagesdf}b) fails to capture them.
However, it is not reflected in IoU values in Fig.~\ref{fig:imagesdf} where grid encoding gave the highest IoUs.
Additionally, local positional encoding has another benefit of faster convergence in the early training phase.
Fig.~\ref{fig:imagesdfconverge} compares grid encoding and local positional encoding with three different training iterations.
Our method converges faster and captures the fine details even in the earlier training stages, such as $64$ and $1,024$ training iterations, compared to grid encoding.

\begin{table*}
    \caption{ Comparisons of IoU with multi-resolution grids for SDFs geometry rendering. "Multi" stands for multi-resolution grid encoding, "Hash" for multi-resolution hash encoding, and "LPE" for local positional encoding. The bold numbers show the best results for each model. }
    \label{tab:multiSdf}
    \centering
    \small
    \begin{tabular}{p{5mm}||c c c|c c c|c c c }
    \toprule 
        & \multicolumn{3}{c|}{\textsc{Armadillo}}  & \multicolumn{3}{c|}{\textsc{Lucy}}  & \multicolumn{3}{c}{\textsc{Thai Statue}} \\
    \midrule
        & Multi & Hash & LPE & Multi & Hash & LPE & Multi & Hash & LPE \\ 
    \midrule
    IoU & 0.9904 & \textbf{0.9941} & 0.9920 & 0.9702 & \textbf{0.9856} & 0.9723 & 0.9712 & \textbf{0.9848} & 0.9736 \\
    \bottomrule
    \end{tabular}
\end{table*}

\begin{figure*}
    \centering
    \setlength{\tabcolsep}{0.002\linewidth}
    \begin{tabular}{ccccc}
        \raisebox{0.11\linewidth}{\rotatebox[origin = c]{90}{\textsc{Thai Statue}}} &
        \begin{overpic}[width=0.24\linewidth]{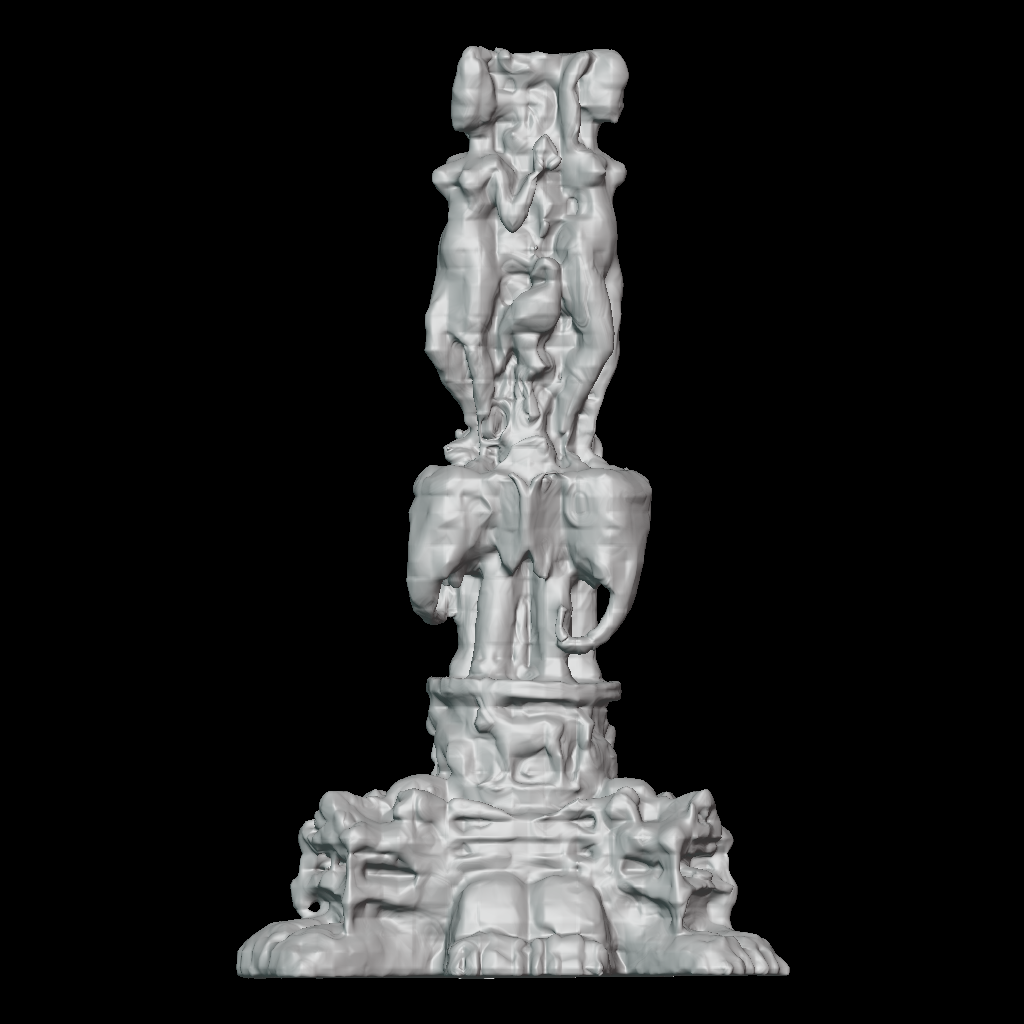}
            \put(443, 659){ \linethickness{0.1mm}\color{red}\polygon(0,0)(99,0)(99,99)(0,99)}
            \put(730, 255){ \includegraphics[width=0.058\linewidth,viewport=474 676 574 776, clip, cfbox=red 0.5pt 0pt]{figs/sdf/cmpMultiRes/MuitiLevel_sdf_GRID_thaiStatue_512_16384_3Level.png} }
            \put(441, 224){ \linethickness{0.1mm}\color{blue}\polygon(0,0)(99,0)(99,99)(0,99)}
            \put(730, 5){ \includegraphics[width=0.058\linewidth,viewport=469 231 569 331, clip, cfbox=blue 0.5pt 0pt]{figs/sdf/cmpMultiRes/MuitiLevel_sdf_GRID_thaiStatue_512_16384_3Level.png} }
        \end{overpic} &
        \begin{overpic}[width=0.24\linewidth]{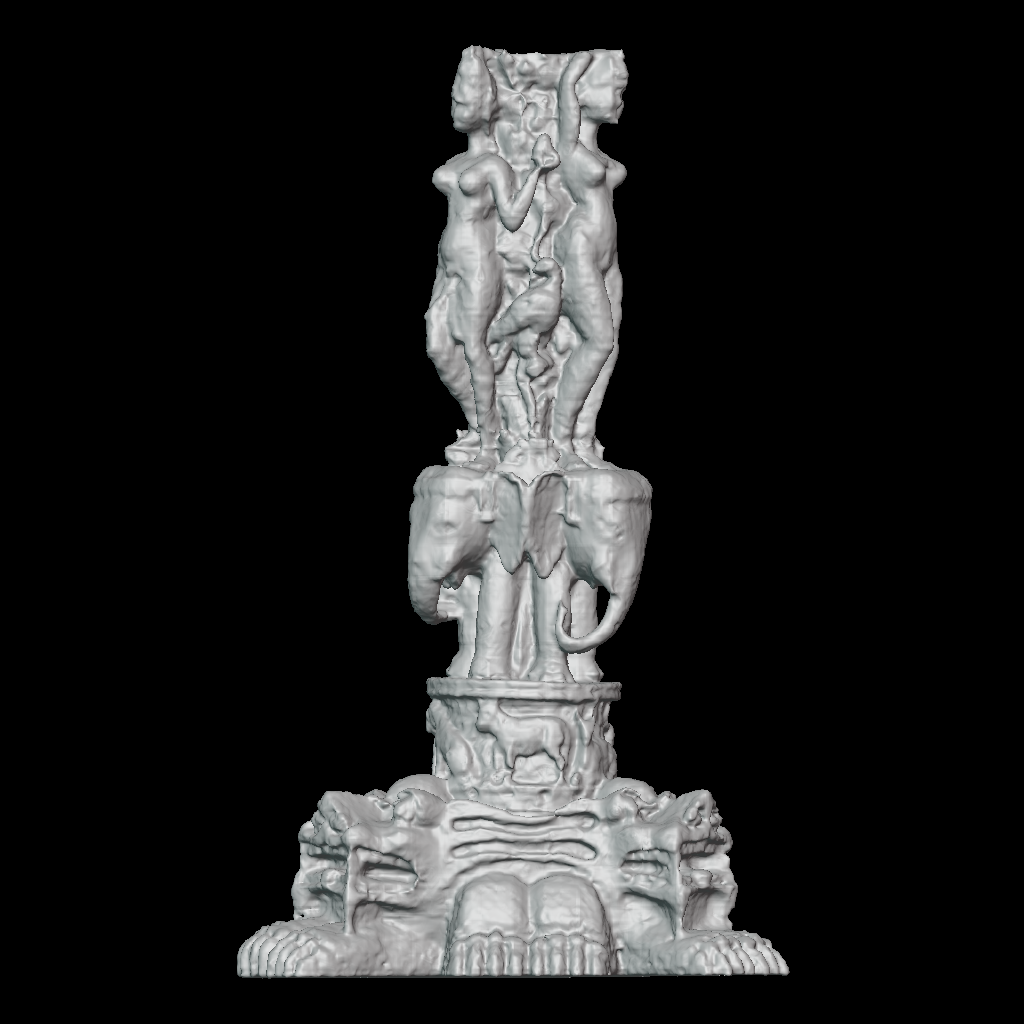} 
            \put(443, 659){ \linethickness{0.1mm}\color{red}\polygon(0,0)(99,0)(99,99)(0,99)}
            \put(730, 255){ \includegraphics[width=0.058\linewidth,viewport=474 676 574 776, clip, cfbox=red 0.5pt 0pt]{figs/sdf/cmpMultiRes/MuitiLevel_sdf_GRID_thaiStatue_512_16384_4Level_hash.png} }
            \put(441, 224){ \linethickness{0.1mm}\color{blue}\polygon(0,0)(99,0)(99,99)(0,99)}
            \put(730, 5){ \includegraphics[width=0.058\linewidth,viewport=469 231 569 331, clip, cfbox=blue 0.5pt 0pt]{figs/sdf/cmpMultiRes/MuitiLevel_sdf_GRID_thaiStatue_512_16384_4Level_hash.png} }
        \end{overpic}&
        \begin{overpic}[width=0.24\linewidth]{figs/sdf/cmp3encs_2x/auto_sdf_LPE_thaiStatue_512_itr16384_A.png} 
            \put(443, 659){ \linethickness{0.1mm}\color{red}\polygon(0,0)(99,0)(99,99)(0,99)}
            \put(730, 255){ \includegraphics[width=0.058\linewidth,viewport=474 676 574 776, clip, cfbox=red 0.5pt 0pt]{figs/sdf/cmp3encs_2x/auto_sdf_LPE_thaiStatue_512_itr16384_A.png} }
            \put(441, 224){ \linethickness{0.1mm}\color{blue}\polygon(0,0)(99,0)(99,99)(0,99)}
            \put(730, 5){ \includegraphics[width=0.058\linewidth,viewport=469 231 569 331, clip, cfbox=blue 0.5pt 0pt]{figs/sdf/cmp3encs_2x/auto_sdf_LPE_thaiStatue_512_itr16384_A.png} }
        \end{overpic} &
        \begin{overpic}[width=0.24\linewidth]{figs/sdf/sdf_thai_512_ref.png} 
            \put(443, 659){ \linethickness{0.1mm}\color{red}\polygon(0,0)(99,0)(99,99)(0,99)}
            \put(730, 255){ \includegraphics[width=0.058\linewidth,viewport=474 676 574 776, clip, cfbox=red 0.5pt 0pt]{figs/sdf/sdf_thai_512_ref.png} }
            \put(441, 224){ \linethickness{0.1mm}\color{blue}\polygon(0,0)(99,0)(99,99)(0,99)}
            \put(730, 5){ \includegraphics[width=0.058\linewidth,viewport=469 231 569 331, clip, cfbox=blue 0.5pt 0pt]{figs/sdf/sdf_thai_512_ref.png} }
        \end{overpic} \\

        & IoU = 0.9712 & \textbf{ IoU = 0.9848 } & IoU = 0.9736 \\
        \addlinespace

        & \small{(a) Multi-resolution Grid Encoding  } & \small{(b) Multi-resolution Hash Encoding} & \small{(c) Local Positional Encoding } & \small{(d) Reference}
    \end{tabular}
    \caption{ Qualitative and quantitative comparison with multi-resolution grid encoding, multi-resolution hash encoding~\cite{10.1145/3528223.3530127}, and local positional encoding with SDFs geometry rendering. The parameters of all the encodings are configured for approximately the same amount of latent vectors for a fair comparison. The bold number represents the best in the encodings for the geometry. }
    \label{fig:imagesdfmultires}
\end{figure*}

\subsection{Comparison with Multi-resolution Grid} \label{sec:cmpMulti}
We compared our local positional encoding only with the single low-resolution grid in Sec.~\ref{sec:image} and Sec.~\ref{sec:sdf}.
However, multi-resolution grids containing higher-resolution grids usually achieve better quality than a single-level grid.
Thus, in this section, we evaluate our encoding against multi-resolution grid encoding and multi-resolution hash encoding~\cite{10.1145/3528223.3530127}, especially for SDFs.

We use almost the same number of trainable parameters in grids for each encoding for a fair comparison.
We choose $3$-level grids ($N = 16, 32, 64$) with $2$-dimensional latent vectors as multi-resolution grid encoding, and $4$-level grids ($N = 16, 32, 64, 128$) with $2$-dimensional latent vectors using $2^{17}$ hash table size for multi-resolution hash encoding.
Note that multi-resolution hash encoding has hash collisions on the higher-level grids due to the fixed-size hash table while multi-resolution grid encoding does not.
The configuration of local positional encoding in this comparison is the same as used in Sec.~\ref{sec:sdf}.
These settings result in roughly $590$k parameters for all the encodings in the grids.
Fig.~\ref{fig:imagesdfmultires} shows the rendered images with these three encodings for the \textsc{Thai Statue} model.
The rendered images for more geometries can be found in the supplemental document.
We can see that our encoding achieved visually better results than multi-resolution grid encoding and comparable results to multi-resolution hash encoding only using a single-level grid.
Also, taking a closer look, multi-resolution hash encoding produces micro-structured artifacts on the smooth surface due to hash collisions, which cannot be seen with our encoding, though higher-resolution grids accepting hash collisions can capture high-frequency details well.
On the other hand, in Table~\ref{tab:multiSdf} showing the quantitative comparisons, multi-resolution hash encoding gave the highest score in the IoU metric while despite using only a single-level grid, local positional encoding achieved a higher IoU value than multi-resolution grid encoding.
Note that though our encoding uses only a single-level grid, it can be extended to multi-resolution grids, with which we expect to improve our encoding further.
It is conceivable to adopt multi-resolution grids in local positional encoding in future work.

\section{Conclusions}
In this paper, we introduced a novel input encoding method for an MLP, local positional encoding, to allow a small MLP to learn high-frequency signals with a less memory footprint.
Our proposed method uses a grid to store the weights (i.e. the latent coefficients) for each sinusoidal function of positional encoding and optimizes them along with the network weights through stochastic gradient descent.
Learning the latent coefficients in each grid cell locally controls the amplitudes of the encodings, so it well adopts the spatially varying signals.
We demonstrated the effectiveness of local positional encoding against positional and grid encodings for the 2D image reconstruction and 3D signed distance functions.
In both tasks, our method can represent better-quality results using a small MLP.
Additionally, our encoding also shows comparable results even in comparisons with multi-resolution grid encodings in the SDFs reconstruction problem.

Although local positional encoding captures high-frequency information well only with a small single-level grid, it produces axis-aligned artifacts.
We believe they are the inheritance of positional encoding which also suffers from such artifacts.
We leave it to future work to investigate an efficient way of alleviating the artifacts.
In addition, applying multi-resolution grids to local positional encoding is a straightforward extension to improve it further.
Also, our method requires storing the latent coefficients for each dimensional input.
Therefore, our encoding in higher dimensional tasks results in more memory intensive.
Reducing the total number of latent coefficients in each grid cell is our interesting future work.

\section*{Acknowledgments}
We thank colleagues in Advanced Rendering Research (ARR) Group for discussion, neural network implementation, and proofreading. The \textsc{Bridge} and \textsc{Fish Market} images are from \href{https://www.pexels.com/photo/red-and-brown-bridge-2932346/}{Anand Dandekar} and \href{https://www.pexels.com/photo/food-shopping-market-fish-10305777/}{Sofía Rabassa}, respectively. We would also like to thank the Stanford Computer Graphics Laboratory for \textsc{Armadillo}, \textsc{Lucy}, and \textsc{Thai Statue} models.

\bibliographystyle{eg-alpha-doi} 
\bibliography{main}

\newcommand{\etalchar}[1]{$^{#1}$}
\begin{thebibliography}{\uppercase{KRWM22}}

\bibitem[{AMD}21]{hip}
\textsc{{AMD}}:
\newblock Hip programming guide v4.5.
\newblock \url{https://rocmdocs.amd.com/en/latest/Programming_Guides/HIP-GUIDE.html}, 2021.

\bibitem[BMT{\etalchar{*}}21]{9710056}
\textsc{Barron J.~T., Mildenhall B., Tancik M., Hedman P., Martin-Brualla R., Srinivasan P.~P.}:
\newblock Mip-nerf: A multiscale representation for anti-aliasing neural radiance fields.
\newblock In \emph{2021 IEEE/CVF International Conference on Computer Vision (ICCV)} (2021), pp.~5835--5844.
\newblock \href {https://doi.org/10.1109/ICCV48922.2021.00580} {\path{doi:10.1109/ICCV48922.2021.00580}}.

\bibitem[BMV{\etalchar{*}}22]{9878829}
\textsc{Barron J.~T., Mildenhall B., Verbin D., Srinivasan P.~P., Hedman P.}:
\newblock Mip-nerf 360: Unbounded anti-aliased neural radiance fields.
\newblock In \emph{2022 IEEE/CVF Conference on Computer Vision and Pattern Recognition (CVPR)} (2022), pp.~5460--5469.
\newblock \href {https://doi.org/10.1109/CVPR52688.2022.00539} {\path{doi:10.1109/CVPR52688.2022.00539}}.

\bibitem[BMV{\etalchar{*}}23]{barron2023zipnerf}
\textsc{Barron J.~T., Mildenhall B., Verbin D., Srinivasan P.~P., Hedman P.}:
\newblock Zip-nerf: Anti-aliased grid-based neural radiance fields.
\newblock \emph{arXiv} (2023).

\bibitem[CAPM20]{chibane20ifnet}
\textsc{Chibane J., Alldieck T., Pons-Moll G.}:
\newblock Implicit functions in feature space for 3d shape reconstruction and completion.
\newblock In \emph{{IEEE} Conference on Computer Vision and Pattern Recognition (CVPR)} (jun 2020), {IEEE}.

\bibitem[GB10]{Glorot2010UnderstandingTD}
\textsc{Glorot X., Bengio Y.}:
\newblock Understanding the difficulty of training deep feedforward neural networks.
\newblock In \emph{International Conference on Artificial Intelligence and Statistics} (2010).

\bibitem[HCZ21]{10.1145/3478513.3480569}
\textsc{Hadadan S., Chen S., Zwicker M.}:
\newblock Neural radiosity.
\newblock \emph{ACM Trans. Graph. 40}, 6 (dec 2021).
\newblock URL: \url{https://doi.org/10.1145/3478513.3480569}, \href {https://doi.org/10.1145/3478513.3480569} {\path{doi:10.1145/3478513.3480569}}.

\bibitem[HPG{\etalchar{*}}21]{NEURIPS2021_4a06d868}
\textsc{Hertz A., Perel O., Giryes R., Sorkine-hornung O., Cohen-or D.}:
\newblock Sape: Spatially-adaptive progressive encoding for neural optimization.
\newblock In \emph{Advances in Neural Information Processing Systems} (2021), Ranzato M., Beygelzimer A., Dauphin Y., Liang P., Vaughan J.~W., (Eds.), vol.~34, Curran Associates, Inc., pp.~8820--8832.
\newblock URL: \url{https://proceedings.neurips.cc/paper_files/paper/2021/file/4a06d868d044c50af0cf9bc82d2fc19f-Paper.pdf}.

\bibitem[KB15]{Adam}
\textsc{Kingma D.~P., Ba J.}:
\newblock Adam: {A} method for stochastic optimization.
\newblock In \emph{3rd International Conference on Learning Representations, {ICLR} 2015, San Diego, CA, USA, May 7-9, 2015, Conference Track Proceedings} (2015), Bengio Y., LeCun Y., (Eds.).

\bibitem[KMX{\etalchar{*}}21]{10.1145/3450626.3459795}
\textsc{Kuznetsov A., Mullia K., Xu Z., Ha\v{s}an M., Ramamoorthi R.}:
\newblock Neumip: Multi-resolution neural materials.
\newblock \emph{ACM Trans. Graph. 40}, 4 (jul 2021).
\newblock URL: \url{https://doi.org/10.1145/3450626.3459795}, \href {https://doi.org/10.1145/3450626.3459795} {\path{doi:10.1145/3450626.3459795}}.

\bibitem[KRWM22]{ReluField_sigg_22}
\textsc{Karnewar A., Ritschel T., Wang O., Mitra N.}:
\newblock Relu fields: The little non-linearity that could.
\newblock In \emph{ACM SIGGRAPH 2022 Conference Proceedings} (New York, NY, USA, 2022), SIGGRAPH '22, Association for Computing Machinery.
\newblock URL: \url{https://doi.org/10.1145/3528233.3530707}, \href {https://doi.org/10.1145/3528233.3530707} {\path{doi:10.1145/3528233.3530707}}.

\bibitem[MESK22]{10.1145/3528223.3530127}
\textsc{M\"{u}ller T., Evans A., Schied C., Keller A.}:
\newblock Instant neural graphics primitives with a multiresolution hash encoding.
\newblock \emph{ACM Trans. Graph. 41}, 4 (jul 2022).
\newblock URL: \url{https://doi.org/10.1145/3528223.3530127}, \href {https://doi.org/10.1145/3528223.3530127} {\path{doi:10.1145/3528223.3530127}}.

\bibitem[MGB{\etalchar{*}}21]{mehta2021modulated}
\textsc{Mehta I., Gharbi M., Barnes C., Shechtman E., Ramamoorthi R., Chandraker M.}:
\newblock Modulated periodic activations for generalizable local functional representations.
\newblock In \emph{Proceedings of the IEEE/CVF International Conference on Computer Vision (ICCV)} (October 2021), pp.~14214--14223.

\bibitem[MLL{\etalchar{*}}21]{martel2021acorn}
\textsc{Martel J.~N., Lindell D.~B., Lin C.~Z., Chan E.~R., Monteiro M., Wetzstein G.}:
\newblock Acorn: Adaptive coordinate networks for neural representation.
\newblock \emph{ACM Trans. Graph. (SIGGRAPH)} (2021).

\bibitem[MST{\etalchar{*}}21]{10.1145/3503250}
\textsc{Mildenhall B., Srinivasan P.~P., Tancik M., Barron J.~T., Ramamoorthi R., Ng R.}:
\newblock Nerf: Representing scenes as neural radiance fields for view synthesis.
\newblock \emph{Commun. ACM 65}, 1 (dec 2021), 99–106.
\newblock URL: \url{https://doi.org/10.1145/3503250}, \href {https://doi.org/10.1145/3503250} {\path{doi:10.1145/3503250}}.

\bibitem[RBA{\etalchar{*}}19]{pmlr-v97-rahaman19a}
\textsc{Rahaman N., Baratin A., Arpit D., Draxler F., Lin M., Hamprecht F., Bengio Y., Courville A.}:
\newblock On the spectral bias of neural networks.
\newblock In \emph{Proceedings of the 36th International Conference on Machine Learning} (09--15 Jun 2019), Chaudhuri K., Salakhutdinov R., (Eds.), vol.~97 of \emph{Proceedings of Machine Learning Research}, PMLR, pp.~5301--5310.
\newblock URL: \url{https://proceedings.mlr.press/v97/rahaman19a.html}.

\bibitem[SMGG01]{LitSphere}
\textsc{Sloan P.-P.~J., Martin W., Gooch A., Gooch B.}:
\newblock The lit sphere: A model for capturing npr shading from art.
\newblock In \emph{Proceedings of Graphics Interface 2001} (CAN, 2001), GI '01, Canadian Information Processing Society, pp.~143--150.

\bibitem[STH{\etalchar{*}}19]{sitzmann2019deepvoxels}
\textsc{Sitzmann V., Thies J., Heide F., Nie{\ss}ner M., Wetzstein G., Zollh{\"o}fer M.}:
\newblock Deepvoxels: Learning persistent 3d feature embeddings.
\newblock In \emph{Proc. Computer Vision and Pattern Recognition (CVPR), IEEE} (2019).

\bibitem[TET{\etalchar{*}}22]{10.1145/3528233.3530727}
\textsc{Takikawa T., Evans A., Tremblay J., M\"{u}ller T., McGuire M., Jacobson A., Fidler S.}:
\newblock Variable bitrate neural fields.
\newblock In \emph{ACM SIGGRAPH 2022 Conference Proceedings} (New York, NY, USA, 2022), SIGGRAPH '22, Association for Computing Machinery.
\newblock URL: \url{https://doi.org/10.1145/3528233.3530727}, \href {https://doi.org/10.1145/3528233.3530727} {\path{doi:10.1145/3528233.3530727}}.

\bibitem[TLY{\etalchar{*}}21]{Takikawa2021NeuralGL}
\textsc{Takikawa T., Litalien J., Yin K., Kreis K., Loop C.~T., Nowrouzezahrai D., Jacobson A., McGuire M., Fidler S.}:
\newblock Neural geometric level of detail: Real-time rendering with implicit 3d shapes.
\newblock \emph{2021 IEEE/CVF Conference on Computer Vision and Pattern Recognition (CVPR)} (2021), 11353--11362.

\bibitem[TSM{\etalchar{*}}20]{tancik2020fourfeat}
\textsc{Tancik M., Srinivasan P.~P., Mildenhall B., Fridovich-Keil S., Raghavan N., Singhal U., Ramamoorthi R., Barron J.~T., Ng R.}:
\newblock Fourier features let networks learn high frequency functions in low dimensional domains.
\newblock \emph{NeurIPS} (2020).

\bibitem[VSW{\etalchar{*}}23]{ntc2023}
\textsc{Vaidyanathan K., Salvi M., Wronski B., Akenine-Möller T., Ebelin P., Lefohn A.}:
\newblock {Random-Access Neural Compression of Material Textures}.
\newblock In \emph{Proceedings of SIGGRAPH} (2023).

\bibitem[WBSS04]{1284395}
\textsc{Wang Z., Bovik A., Sheikh H., Simoncelli E.}:
\newblock Image quality assessment: from error visibility to structural similarity.
\newblock \emph{IEEE Transactions on Image Processing 13}, 4 (2004), 600--612.
\newblock \href {https://doi.org/10.1109/TIP.2003.819861} {\path{doi:10.1109/TIP.2003.819861}}.

\bibitem[WZK{\etalchar{*}}23]{Weier:2023}
\textsc{Weier P., Zirr T., Kaplanyan A., Yan L.-Q., Slusallek P.}:
\newblock Neural prefiltering for correlation-aware levels of detail.
\newblock In \emph{Proceedings of SIGGRAPH 2023} (2023).

\end{thebibliography}

\end{document}